\documentclass[nonacm,sigplan]{acmart}

\pdfoutput=1
\usepackage{hyperref}   % Load first
\usepackage{hyperxmp}   % Load after hyperref
\usepackage{booktabs}
\usepackage{subcaption} % Add this in your preamble if not already included
\usepackage{enumitem}
\usepackage{algorithm}
\usepackage[noend]{algpseudocode}
\usepackage{makecell}
\usepackage{multicol}
\usepackage{multirow}

\usepackage{setspace}

\usepackage{etoolbox}
\usepackage{tcolorbox}

\newcommand{\THISWORK}{{\fontfamily{lmss}\selectfont
MoDM}}

\begin{document}

%\title{\THISWORK: Efficient Diffusion Model Serving Using Mixture-of-Models}
\title{\THISWORK: Efficient Serving for Image Generation via \underline{M}ixture-\underline{o}f-\underline{D}iffusion \underline{M}odels}
\author{Yuchen Xia$^1$, Divyam Sharma$^1$, Yichao Yuan$^1$, Souvik Kundu$^2$, and Nishil Talati$^1$\\
$^1$University of Michigan, USA; $^2$Intel Labs, USA}
% removed for anonymity

% \begin{abstract}
% Please look at the ASPLOS Call For Papers for formatting instructions.

% ESPLOS, the Espresso-Powered Latte Operating System, is an innovative platform that combines an operating system (OS), a domain-specific programming language (PL), and a sophisticated compiler to consistently brew perfect espresso shots and create stunning latte art. This paper introduces ESPLOS and presents simulation results, evaluation data, and user studies, highlighting its significant advancements in coffee machine technology.
% \end{abstract}
% \input{asplos25-templates/contribution}
\begin{abstract}
Diffusion-based text-to-image generation models trade latency for quality: small models are fast but generate lower quality images, while large models produce better images but are slow.
We present \THISWORK, a novel caching-based serving system for diffusion models that \textbf{dynamically} balances latency and quality through a \textit{mixture of diffusion models}.
Unlike prior approaches that rely on model-specific internal features, \THISWORK\ caches \textit{final images}, allowing seamless retrieval and reuse across multiple diffusion model families.
This design enables adaptive serving by \textit{dynamically balancing latency and image quality}: using smaller models for cache-hit requests to reduce latency while reserving larger models for cache-miss requests to maintain quality.
Small model image quality is preserved using retrieved cached images.
We design a global monitor that optimally allocates GPU resources and balances inference workload, ensuring high throughput while meeting Service-Level Objectives (SLOs) under varying request rates.
Our evaluations show that \THISWORK\ significantly reduces an average serving time by 2.5$\times$ while retaining image quality, making it a practical solution for scalable and resource-efficient model deployment.
\end{abstract}

\maketitle % should come after the abstract
\pagestyle{plain} % should come right after \maketitle

\section{Introduction} \label{section:introduction}

\begin{figure}[t]
    \centering
    \includegraphics[width=\linewidth]{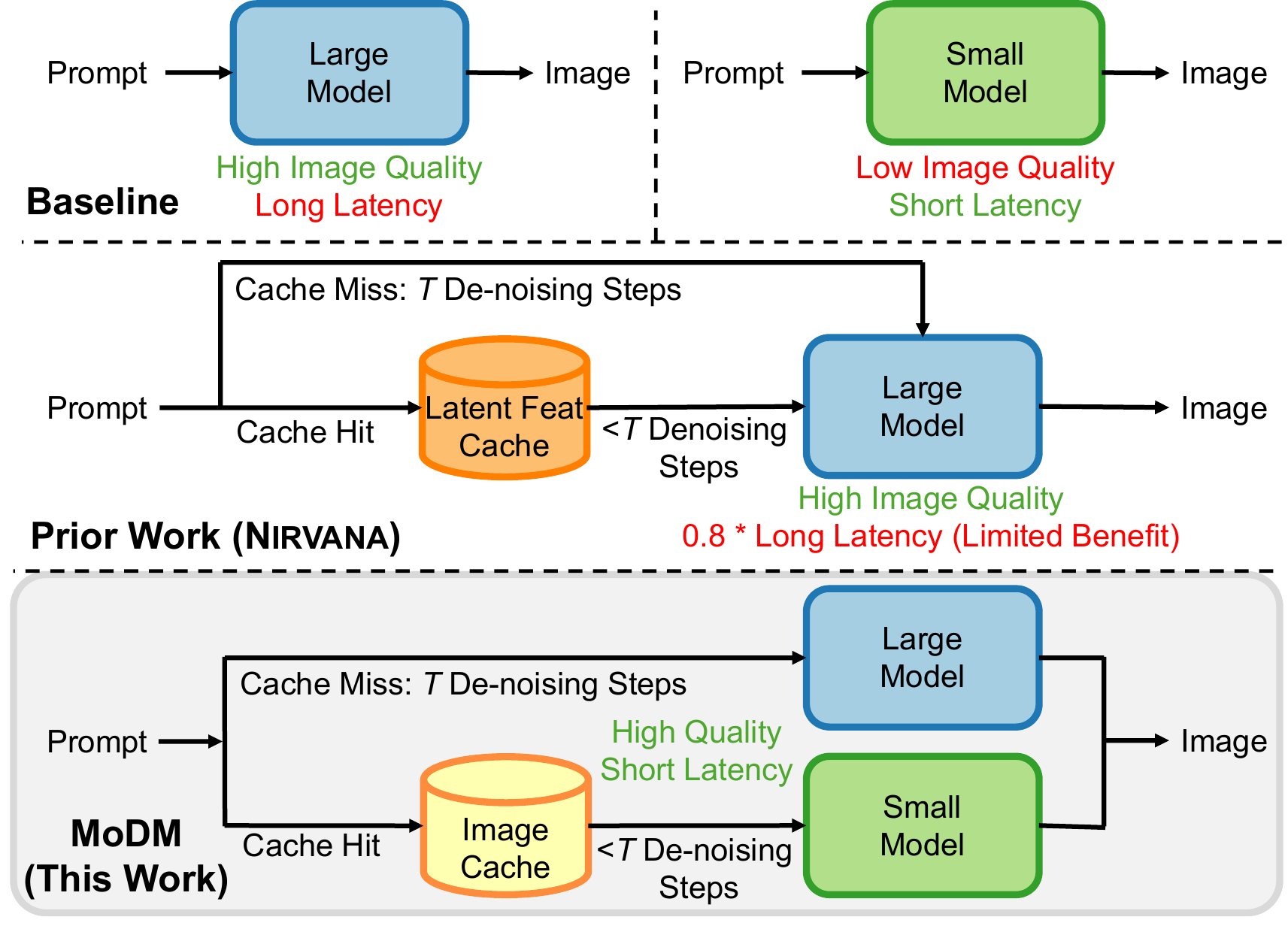}
    \vspace{-0.7cm}
    \caption{Overview of \THISWORK\ that balances between high image quality and short latency using a mixture of models.}
    \vspace{-0.6cm}
    \label{fig:overview}
\end{figure}

% \begin{itemize}
%     \item \textbf{Motivation:}
%     \begin{itemize}
%         \item Text-to-image generation models are resource-intensive and latency-bound.
%         \item Existing systems like NIRVANA rely on latent caches, which have high storage requirements.
%         \item A need for a more scalable and efficient system architecture.
%     \end{itemize}
%     \item \textbf{Contributions:}
%     \begin{itemize}
%         \item Propose replacing latent caches with image caches for smaller storage footprints.
%         \item Optimize workload allocation by offloading cache-hit requests to smaller models, ensuring lower latency and higher system throughput.
%         \item Employing a dynamic model loading policy to adapt to changing request rates with minimal sacrifice in image quality. 
%     \end{itemize}
% \end{itemize}

Diffusion models have revolutionized text-to-image generation, enabling the creation of high-quality, photorealistic images from natural language descriptions.
Their success has made them a cornerstone of AI-driven creative tools, powering applications in digital art, content generation, and interactive media.
The demand for diffusion models is at an all-time high.
Adobe’s Firefly service~\cite{adobe-announcement} has generated over 2 billion images, while OpenAI’s DALL-E 2~\cite{openai-dalle2} has seen similar adoption, alongside an exponential rise in prompt submissions to Stable Diffusion systems, as shown by DiffusionDB~\cite{wang2022diffusiondb}.
However, each diffusion model inference takes 10s of seconds, making it a computationally expensive task.
Meeting this growing demand requires significant improvements in serving system throughput and latency.

An effective way to improve diffusion model performance is through \textit{caching}, which reduces redundant computations and accelerates inference.
Prior works have explored various caching techniques~\cite{blockCache, patchedServe, nirvama, deepCache, layerCaching, fora, dualCaching, ReCon}, including latent caching (storing intermediate noisy features), feature caching (storing activations), and image patch caching.
While these techniques enhance performance, caching intermediate features has two major limitations.
First, it restricts serving to a single model, as cached content is model-specific. 
Second, relying on just one model limits the potential performance gains, reducing flexibility in optimizing inference.
% While these techniques improve performance, we identify that caching intermediate features has two significant limitations: (1) it limits the approach to using only one model as cached content is model-specific, and (2) using only one model limits the amount of benefits.
% they introduce challenges such as model-specific constraints, high storage overhead, and limited cross-model compatibility.
% Moreover, caching intermediate representations often lacks semantic awareness, making retrieval less effective for diverse prompts.
% Addressing these challenges, \THISWORK\ instead caches final generated images, enabling a universal caching strategy that supports multiple diffusion models while leveraging CLIP embeddings for robust text-image similarity matching.

In this paper, we introduce \THISWORK, an efficient serving system for text-to-image generation using a \textit{mixture of diffusion models}.
While our work is based on the general idea of caching, \THISWORK\ is uniquely designed to achieve three design goals.
(1) Diffusion models trade off inference latency and image quality: large models generate high-quality images but are slow, while smaller models are faster but sacrifice quality.
\THISWORK\ dynamically exploits this trade-off by \textit{designing a caching strategy} compatible with multiple models to optimize both speed and quality.
(2) To maintain high-quality image generation, \THISWORK\ implements an \textit{advanced retrieval strategy} that ensures retrieved cached content closely aligns with new prompts.
(3) Finally, \THISWORK\ introduces an adaptive serving system that \textit{dynamically balances latency and quality} based on request rates and system load, ensuring efficient and scalable performance.

Fig.~\ref{fig:overview} shows the design overview of \THISWORK.
To ensure that the cache content is accessible and relevant for multiple models, we propose to \textit{cache final generated images} in the past, in contrast to caching latent intermediate images in prior work~\cite{nirvama}.
When a new prompt closely matches a cached image, \THISWORK\ retrieves the image, applies controlled noise, and refines it using a smaller model.
This approach preserves the quality of the cached image while benefiting from the lower latency of the smaller model.
Requests that miss the cache are processed using a large model.
This hybrid approach effectively balances latency and image quality by using a \textit{mixture of models}.

Caching final images enables retrieval based on text-to-image similarity, unlike prior works~\cite{nirvama} that rely solely on text-to-text similarity by caching intermediate features.  
Leveraging the CLIP score~\cite{clipScore} and image generation examples, we demonstrate that text-to-image similarity retrieval better aligns with user prompts, using it for cache retrieval in \THISWORK.
Building on this, \THISWORK\ integrates image caching and a mixture of models into a high-performance diffusion model serving system.  
The system features a \textit{Request Scheduler} that manages incoming requests, categorizes them into cache hits and misses, retrieves cached images, and maintains cache content over time.
Additionally, a \textit{Global Monitor} analyzes request rates and cache hit/miss distributions to dynamically allocate GPU resources, scheduling different models for inference based on workload conditions.

We evaluate the effectiveness of \THISWORK\ using both performance (\textit{i.e.,} throughput and tail latency) and image quality  (\textit{i.e.,} CLIP~\cite{clipScore} and FID~\cite{FID} scores) metrics.  
Using the DiffusionDB~\cite{wang2022diffusiondb} dataset, we demonstrate that \THISWORK\ achieves a 2.5$\times$ improvement in inference throughput and 46.7\% lower energy consumption, compared to using only a high-quality model, by leveraging Stable Diffusion-3.5-Large as a high-quality model and Stable Diffusion-XL as a low-latency model.
Additionally, we evaluate tail latency under varying request rates, showing that \THISWORK\ sustains significantly higher loads without violating Service Level Objectives (SLOs), outperforming state-of-the-art solutions.
Finally, we highlight the versatility of \THISWORK\ by serving requests across different model families, including Stable Diffusion~\cite{sd3} and SANA~\cite{xie2024sana}.
Unlike SANA that statically reduce inference cost by designing smaller models, \THISWORK\ introduces a novel technique that can intelligently and \textbf{\textit{dynamically} balance latency and quality by leveraging a mixture of diffusion models}.
% Unlike prior algorithmic works like SANA that statically reduce cost by designing smaller models, \THISWORK\ system design \textbf{\textit{dynamically balances between latency and quality}} using a combination of models.
The contributions of \THISWORK\ are as follows.
\begin{itemize}[leftmargin=*]
    \item An optimized cache design and retrieval policy based on final images to accelerate diffusion model inference.
    \item Generating images by retrieving a cached image, adding noise, and de-noising it with a low-cost model.
    \item A hybrid serving approach that leverages small and large models to balance latency and image quality.
    \item \THISWORK: an end-to-end text-to-image serving system design that dynamically adjusts to load variations, achieving 2.5$\times$ performance improvement.
\end{itemize}

\section{Background on Diffusion Model Serving} \label{section:Background}

% This section provides a brief overview of diffusion models, their design trade-offs, and prior work leveraging prompt similarity for optimization.

\subsection{Diffusion Models} \label{sec:background_diffusion}
Diffusion models~\cite{rombach2021highresolution, saharia2022photorealistic, xie2024sana} generate text-prompted images from noise via an iterative de-noising mechanism.
These models employ a forward process adds noise over \( T \) steps, and a reverse process de-noises the image iteratively.
Each de-noising step involves passing the latent representation through the full model, which is computationally expensive, with a typical inference requiring 50 steps~\cite{nirvama}.

This makes diffusion models slower than non-iterative models like GAN~\cite{goodfellow2014generative} or VAE~\cite{kingma2013auto}, presenting challenges for real-time and high-throughput applications.
Diffusion models are evaluated using several metrics: the Frechet Inception Distance (FID) score~\cite{FID}, which measures image realism and fidelity by comparing generated images to real ones; the CLIPScore~\cite{clipScore}, which quantifies the alignment between generated images and their corresponding text prompts; the PickScore~\cite{kirstain2023pick}, which leverages a preference-tuned language–vision model to better reflect human judgment of image–text compatibility; and the Inception Score (IS)~\cite{szegedy2016rethinking}, which evaluates the quality and diversity of generated images based on the confidence and entropy of class predictions from an Inception network.
Diffusion models vary in size, with larger models like Stable Diffusion 3.5-Large~\cite{sd3} and Imagen 3~\cite{baldridge2024imagen} offering high-quality, detailed images at the cost of longer inference times, while smaller models like Stable Diffusion XL~\cite{sdxl} and SANA~\cite{xie2024sana} prioritize speed but sacrifice image fidelity.
While previous studies suggest that high-quality image generation requires a large model~\cite{sd3,nirvama}, our work challenges this by using a mixture of small and large models to optimally balance latency and quality.

\subsection{Caching to Improve Serving Performance} \label{sec:background_caching}
To reduce the computational overhead of iterative de-noising in diffusion models, several prior works~\cite{blockCache, patchedServe, nirvama, deepCache, layerCaching, fora, dualCaching, ReCon} optimize inference based on prompt similarities.
One notable approach, \textsc{Nirvana}~\cite{nirvama}, introduces an approximate caching mechanism that stores intermediate latent representations from previous images.
When a new prompt closely matches a cached prompt, \textsc{Nirvana} retrieves the cached representation and skips several de-noising steps, improving latency.
The system caches multiple latent representations to enable flexibility in retrieval, using larger values of \( k \) for highly similar prompts.
\textsc{Nirvana} uses text-to-text similarity to guide the retrieval process, comparing the current prompt’s text embedding with cached ones to select the most relevant intermediate representation.
This approach reduces inference latency by up to 20\% while maintaining image quality.

\section{Challenges of Mixture-of-Models Design} \label{section:motivation}

% To effectively navigate the latency-quality trade-off (\S\ref{sec:background_latency_quality_tradeoff}), this section explores the challenges and opportunities of using a mixture of diffusion models.
% \SK{why we are suddenly talking about DiffMoM? iut has no earlier context.. it should not be introed here. You first talk about only challenges.}
% In particular, we discuss the following research questions to enable a serving system that employs multiple models for balancing this trade-off using the concept of caching.
% At a high level, \THISWORK\ leverages the concept of caching to enhance performance, addressing the following key research questions.
% To navigate the latency-quality trade-off effectively (\S\ref{sec:background_diffusion}), this section examines the challenges and opportunities of using a mixture of diffusion models. 
% Specifically, we explore key research questions to design a serving system that balances this trade-off through caching (\S\ref{sec:background_caching}).
This section explores the challenges and opportunities of using a model mixture to balance the latency-quality trade-off and addresses key research questions for designing an effective caching-based serving system.
(1) How can we design a cache that minimizes space usage while being model agnostic?
(2) How can we efficiently retrieve cached items to ensure optimal quality of image generation?
(3) How can we best balance high image generation quality with low inference latency?

\subsection{What to Cache?} \label{sec:what_to_cache}

Prior work~\cite{nirvama} has used latent caching, storing multiple intermediate representations to speed up diffusion model inference.
However, it has two main drawbacks: significant storage overhead, with 2.5MB per image due to the need to store multiple latent intermediates (using Stable Diffusion-3.5-Large as an example), compared to 1.4MB for storing only the final image; and model dependence, as latents from one model are incompatible with other models, leading to cache fragmentation and scalability issues in multi-model environments.
An alternative is to cache \textit{full images}, which are universally interpretable and model-independent, simplifying cache management and reducing storage costs.
This approach eliminates the need for separate latent caches, utilizes compressed formats like PNG and JPEG, and allows the dynamic reintroduction of noise to reconstruct intermediate states, enabling compatibility across different models and improving scalability in large-scale serving systems.
\begin{tcolorbox}[width=0.48\textwidth]
\textbf{\textit{Insight:}} \textit{Caching full images reduces storage overhead, eliminates model dependency, and enables broad reuse across different diffusion models.}
\end{tcolorbox}

\begin{figure}[t]
    \centering
    \includegraphics[width=\linewidth]{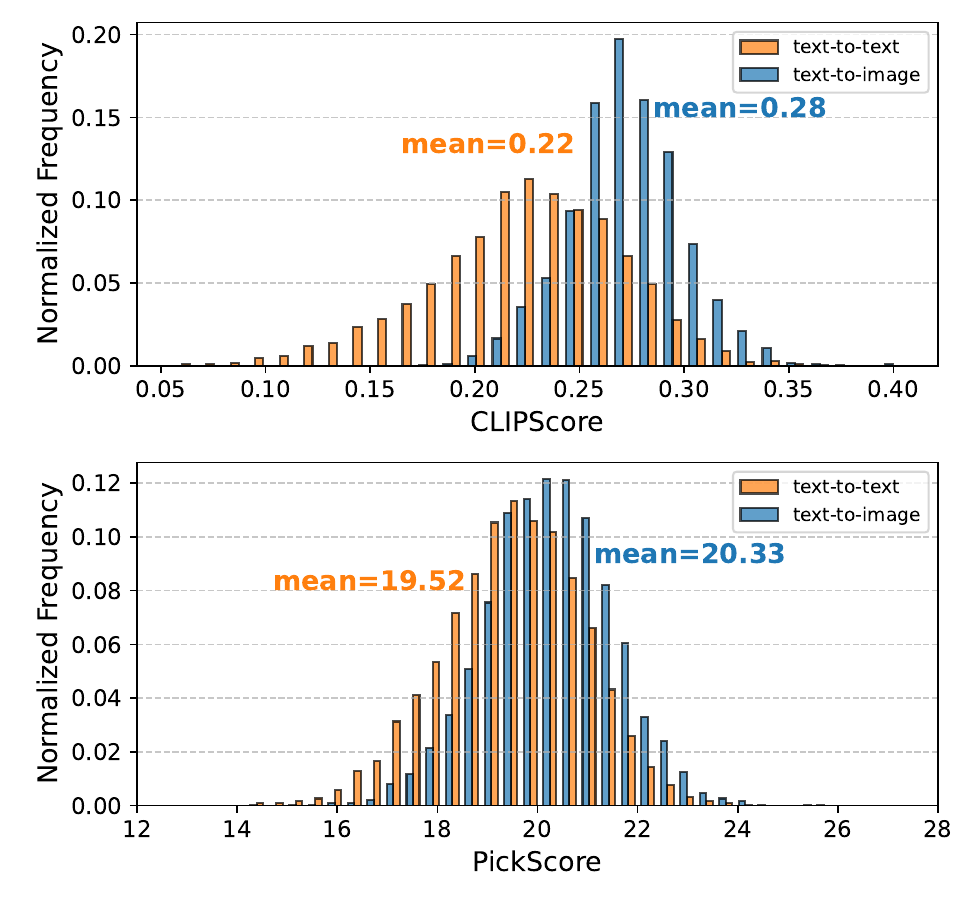} % Replace with actual filename
    \vspace{-0.8cm}
    \caption{Comparison of CLIPScore and PickScore distributions for retrievals based on text-to-text and text-to-image similarity. Higher is better for both scores.}
    \label{fig:retrieval_comparison}
\end{figure}

\subsection{How to Retrieve Cached Items?} \label{sec:how_to_retrieve}
Existing caching methods~\cite{nirvama} use text-to-text similarity for cache retrieval, which often leads to incorrect matches due to a lack of visual alignment.
These methods focus on text embeddings, which do not guarantee that the retrieved image accurately reflects the user’s intent.
Additionally, since prior approaches cache only latent representations rather than full images, they cannot leverage text-to-image similarity, further reducing retrieval precision.
% Existing caching methods~\cite{nirvama} rely on text-to-text similarity for cache retrieval, which often leads to incorrect matches due to a lack of visual alignment.
% These methods focus on semantic closeness in text embeddings, which does not guarantee that the retrieved image accurately reflects the user’s intent.
% Moreover, since prior approaches cache only latent representations rather than full images, they are unable to leverage direct text-to-image similarity for retrieval, further reducing precision.
% As a result, text-based retrieval can produce significant mismatch, limiting the effectiveness of the cache.

\begin{figure}[t]
    \centering
    \includegraphics[width=\linewidth]{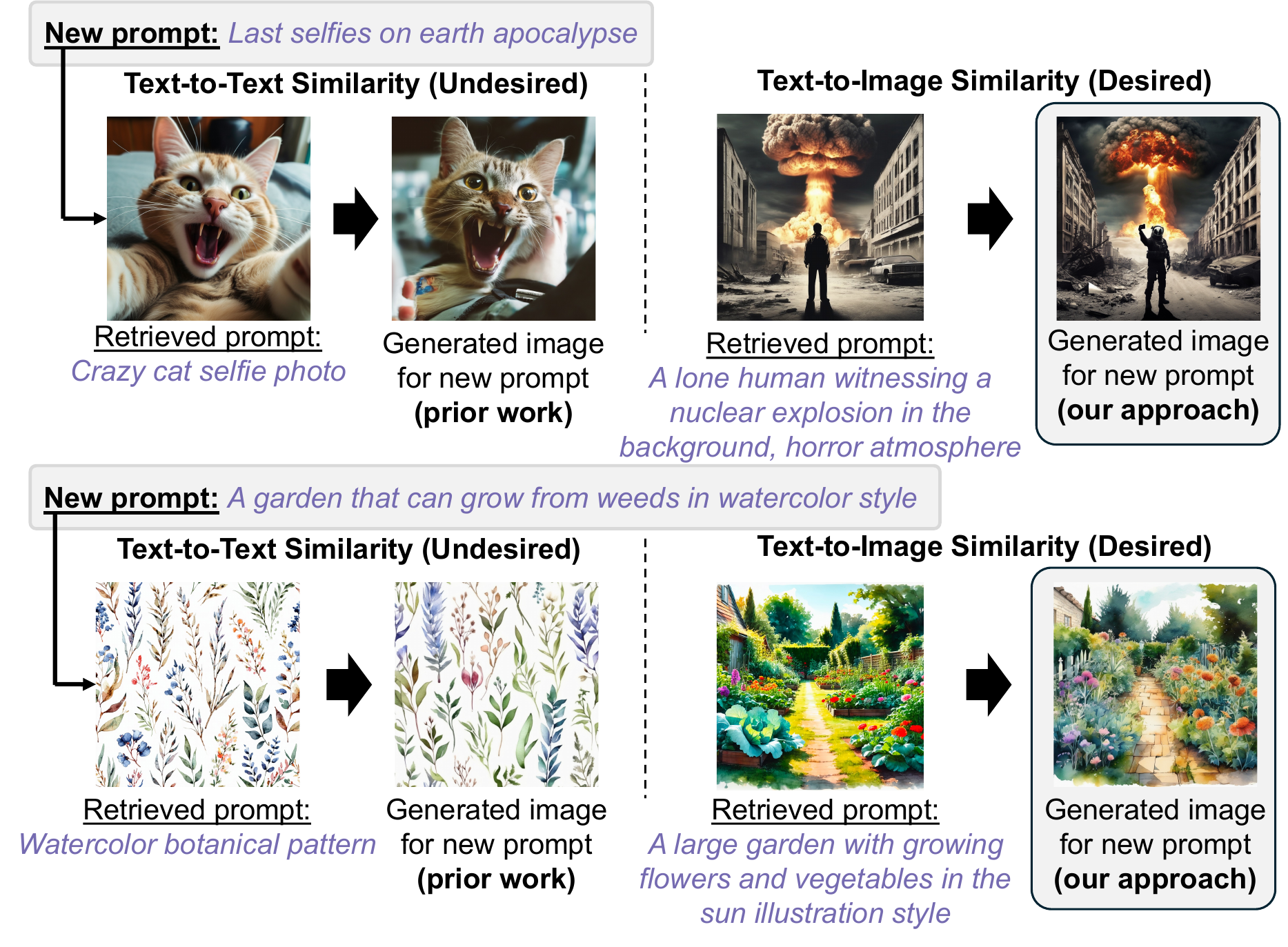} % Replace with actual filename
    \vspace{-0.8cm}
    \caption{Comparison of image quality using a cached image retrieved through text-to-text and text-to-image similarity.}
    \vspace{-0.6cm}
    \label{fig:cache_hit_examples}
\end{figure}
Image caching enables retrieval based on text-to-image similarity, significantly improving alignment with the user’s request in terms of style, structure, and content.
By using CLIP embeddings or similar cross-modal techniques, the system ensures better visual relevance.
As shown in Fig.~\ref{fig:retrieval_comparison}, text-to-image retrieval results in higher CLIPScore compared to text-to-text retrieval, indicating stronger visual alignment between the retrieved image and the new request text. 
To address potential bias from using CLIP for both retrieval and evaluation, we also report PickScore, which similarly favors text-to-image retrieval. Fig.~\ref{fig:cache_hit_examples} highlights cases where text similarity does not align well visually, underscoring the importance of cross-modal retrieval for effective caching.
\begin{tcolorbox}[width=0.48\textwidth]
\textbf{\textit{Insight:}} \textit{Text-to-image similarity-based retrieval is superior to text-to-text similarity because it ensures better visual alignment with user intent.}
\end{tcolorbox}

\subsection{How to Balance Between Latency and Quality?} \label{sec:background_balance_latency_quality}

% Using a single diffusion model for inference is inadequate to effectively balance the trade-off between latency and quality.
% In \textsc{Nirvana}, despite a large cache of 1.5 million latents and a cache hit rate exceeding 90\%, overall computation savings remain at only 20\%, leaving the system vulnerable to high request bursts and frequent SLO violations (more details in \S\ref{section:results}).
% To address this, cache-hit requests can be offloaded to a smaller diffusion model, as recent research suggests that minor refinements do not require a full-scale model~\cite{li2023adaptive, yang2023denoising}.
% Studies~\cite{structured_diffusion, early_structure_late_refinement} on diffusion dynamics further show that early de-noising steps determine structure, while later steps focus on fine details, allowing a lightweight model to efficiently process cache-hit requests with minimal quality loss.
% Performance comparisons between large and small models demonstrate that selectively using a smaller model can significantly reduce computation time while maintaining acceptable output quality, making adaptive model selection a promising approach for real-world serving systems.
% In the following discussion, a large model refers to one comparable in size to Stable Diffusion 3.5, while a small model refers to one similar to Stable Diffusion Tiny.
Using a single diffusion model for inference fails to effectively balance the latency-quality trade-off.
In \textsc{Nirvana}, despite a large cache of 1.5 million latents and a cache hit rate over 90\%, the system only achieves a 20\% reduction in computation, remaining vulnerable to high request bursts and frequent SLO violations (more details in \S\ref{sec:slo_compliance}).
To address this, cache-hit requests can be offloaded to a smaller diffusion model, as recent studies suggest that minor refinements can be efficiently handled by lightweight models~\cite{li2023adaptive, yang2023denoising}.
Research on diffusion dynamics~\cite{structured_diffusion, early_structure_late_refinement} shows that early de-noising steps determine structure, while later steps focus on fine details, enabling smaller models to handle cache-hit requests with minimal quality loss.
Comparing large and small models reveals that adaptive model selection can reduce computation time while maintaining acceptable quality, offering a promising approach for real-world serving systems. 
Here, a large model refers to one like Stable Diffusion-3.5-large, while a small model refers to one like SANA.
\begin{tcolorbox}[width=0.48\textwidth]
\textbf{\textit{Insight:}} \textit{Using a mixture of small and large diffusion models optimizes the latency-quality trade-off, ensuring efficient computation while maintaining high image quality.}
\end{tcolorbox}

\section{\THISWORK\ Design Overview}

% \textcolor{red}{NT: This section should start with design goals - these are the statements that mirror your motivation section, then we should talk about an overview without going into any details, and then design details follow.}

% Our system is designed to optimize the efficiency of serving text-to-image diffusion models by leveraging image caching instead of latent caching and introducing an adaptive model selection strategy. Unlike prior approaches such as NIRVANA, which rely on approximate latent caching, our system stores and retrieves final generated images, enabling cross-model compatibility, significantly reducing storage overhead, and improving the relevance of retrieved images by leveraging text-to-image similarity retrieval instead of text-to-text similarity. Additionally, cache-hit requests are intelligently offloaded to a smaller diffusion model, reducing latency and improving throughput with minimal impact on image quality.

% ========================

This section provides the design goals and overview of the \THISWORK\ serving system.
\vspace{-0.25cm}

\subsection{Design Goals} \label{sec:design_goals}
\THISWORK\ addresses the challenges of serving diffusion model inference workload with the following \textbf{\underline{D}}esign \textbf{\underline{G}}oals (\textbf{DG}).

\textbf{DG\#1: Dynamically Balancing Latency and Quality.}
Building on the trade-off discussed in \S\ref{sec:background_diffusion}, our goal is to achieve inference latency close to a small diffusion model while preserving the image quality of a large model.
\THISWORK\ accomplishes this by strategically leveraging both small and large models for serving.
Unlike prior works that either statically reduce inference costs by designing new models~\cite{xie2024sana} or achieve limited gains by skipping a few de-noising steps~\cite{nirvama}, \THISWORK\ strives to \textit{dynamically} optimize the latency-quality trade-off.
% Rather than using both models for a single request, incoming requests are categorized, ensuring each is served by the most suitable model to optimize efficiency and quality.

\textbf{DG\#2: Compatibility with Multiple Models.}
Different diffusion models and model families (\textit{e.g.,} Stable Diffusion and SANA) offer distinct trade-offs in terms of latency, quality, and architectural optimizations.
Relying on a single model family limits flexibility and may lead to suboptimal performance across diverse request rates.
Our design goal is to enable seamless serving across multiple model families, allowing the system to dynamically select the best model for different requests based on latency considerations.

\textbf{DG\#3: Optimized Cache Design and Retrieval.}
Our caching mechanism is designed to maximize efficiency while maintaining broad compatibility.
The key goals include (1) achieving a high cache hit rate with minimal storage overhead, (2) ensuring cached items remain usable across multiple models (\textbf{DG\#2}), (3) implementing cache management to prevent over-representation of specific items in future generations, and (4) adopting an accurate retrieval that effectively matches new prompts with cached items.

\textbf{DG\#4: Adaptability to Varying Request Rates.}
Request rates fluctuate over time \cite{nirvama, wang2022diffusiondb}, posing challenges for maintaining efficient serving performance. 
Our system is designed to dynamically adapt to these variations by selectively utilizing different models, ensuring an optimal balance between quality and latency under varying workloads.

\subsection{System Design Overview} \label{sec:system_overview}
\begin{figure}[t]
    \centering
    \includegraphics[width=1.0\linewidth]{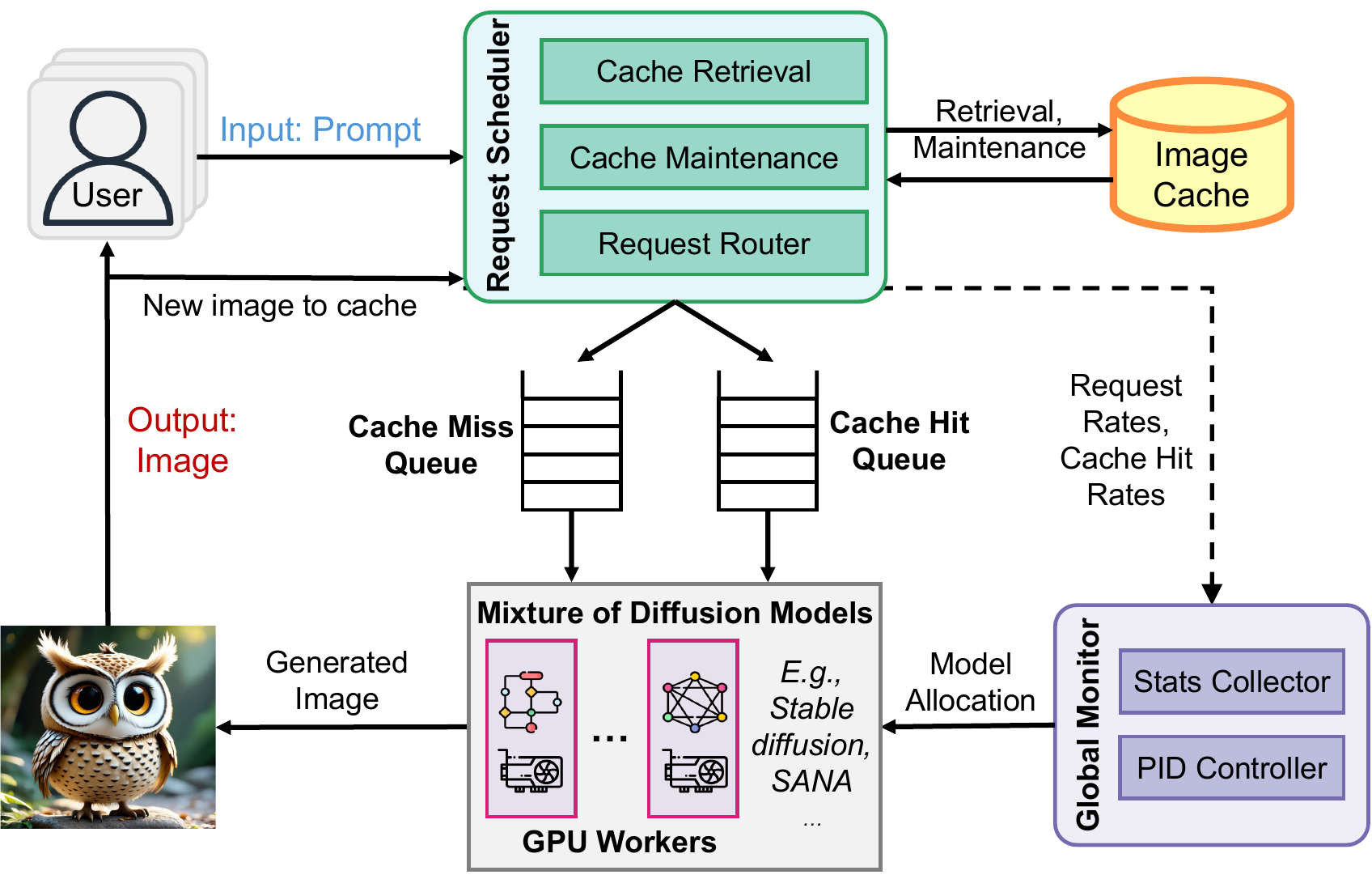}
    \caption{Overview of \THISWORK\ system design.}
    \vspace{-0.8cm}
    \label{fig:sys_overview}
\end{figure}

Fig.~\ref{fig:sys_overview} presents an overview of the \THISWORK\ system architecture, highlighting its key design components.
The system takes a text-based user prompt as \textbf{input} and generates an image as \textbf{output}.
To optimize real-time text-to-image generation with diffusion models, our approach integrates cache-aware request scheduling and dynamic model selection.

The \textbf{Request Scheduler} manages the flow of incoming image generation requests, ensuring efficient resource utilization.
Upon receiving a request, it generates a text embedding using a CLIP model hosted by the scheduler.
This embedding is then analyzed to determine if a cached image closely matches the request, as detailed in \S\ref{sec:cache_retrieval}.
If a sufficiently close match is found, the scheduler retrieves the cached image, enabling content reuse and reducing computational overhead.
For cache-hit requests, the scheduler dynamically adjusts the number of de-noising steps based on the text-to-image similarity score, resuming the generation process from a later stage with lower computation.
If no suitable cached image is identified, the request is marked as a cache miss and undergoes full inference.

To meet \textbf{DG\#1} and optimize throughput, an underlying design principle in \THISWORK\ is to \textbf{serve cache-hit requests with a small model} (more details in \S\ref{sec:generation_using_cache}), which \textit{reduces latency while maintaining the quality of the generated image through cached content}.
\textbf{Cache-miss requests are handled by a larger model}, ensuring high image \textit{generation quality} with full inference from scratch.
The scheduler directs each request to the appropriate queue: cache-hit requests go to the \textit{cache hit queue}, where a refined image is generated with fewer diffusion steps, while cache-miss requests are sent to the \textit{cache miss queue} for full synthesis.
Additionally, to meet \textbf{DG\#3}, the scheduler manages the cache by storing newly generated images and removing older ones, ensuring efficient memory usage and maintaining content diversity.
% To enable fast cache lookup, image embeddings are stored in memory, eliminating the need for recomputation with every request.

The \textbf{Global Monitor} continuously tracks request patterns and system load in real time, leveraging data from the request scheduler to optimize model allocation across GPU workers.
To meet \textbf{DG\#4}, a \textit{PID (Proportional-Integral-Derivative) controller} dynamically adjusts GPU allocation based on key metrics such as request rate, cache hit rate, and de-noising step distribution.
The system strategically balances between model variants, with larger models handling full inference for high-quality outputs and refinements, while smaller models focus solely on efficient refinements of cached images.
By adapting to workload fluctuations, the global monitor ensures optimal resource distribution, preventing resource underutilization while maintaining quality.
These design details are discussed in \S\ref{sec:global_monitor}.

The \textbf{Workers} execute image generation tasks in parallel across multiple GPUs, each hosting a model variant based on the global monitor’s allocation and dynamically switching models as needed.
While workers running larger models are capable of handling both cache-hit and cache-miss requests, they prioritize cache-miss requests, as these require full inference across all denoising steps and therefore incur significantly higher latency.
% \DS{[this sentence was pointed out in nits as confusing hence the changes]}
In contrast, workers running smaller models exclusively process cache-hit requests, focusing on efficient refinement to minimize system-wide latency. 
This dynamic task assignment and model switching strategy ensures optimal GPU utilization, maintaining high throughput while balancing the image generation quality to meet \textbf{DG\#1} and \textbf{DG\#2}.

\section{\THISWORK\ Design Details} \label{section:hardware_design}

In this section, we present the detailed design of \THISWORK, addressing four key research questions.
(1)  How to use cached images for generation by applying a subset of de-noising steps with a small model?
(2) How to effectively retrieve items from the cache and determine the optimal number of de-noising steps for a given prompt?
(3) What is the role of the global monitor in managing GPU resources and model allocation to optimize efficiency under varying load conditions?
(4) How does the request scheduler manage the image cache to ensure efficient system performance?
The notations used in this section are explained in Table~\ref{tab:notation}.

\begin{table}[t]
    \centering
    \scriptsize
    \caption{Notations used in this design.}
    \vspace{-0.4cm}
    \label{tab:notation}
    \begin{tabular}{c l}
        \toprule
        \textbf{Symbol} & \textbf{Description} \\
        \midrule
        \( H_{\text{cache}} \) & Cache hit rate in the last monitoring period. \\
        \( T \) & Total de-noising steps \\ 
        \( K \) & Number of de-noising steps skipped \\ 
        \( P(K = k) \) & Fraction of cache-hit requests assigned to refinement step \( k \) \\
        \( R \) & Recorded request rate in the last monitoring period \\ 
        & (requests per minute). \\
        \( W_{\text{hit}} \) & Cache-hit workload. \\
        \( W_{\text{miss}} \) & Cache-miss workload. \\
        \( N \) & Total number of GPU workers. \\
        \( P_{\text{small}} \) & Profiled throughput of the small model \\ 
        & (requests per minute per GPU). \\
        \( P_{\text{large}} \) & Profiled throughput of the large model \\ 
        & (requests per minute per GPU). \\
        \( N_{\text{small}} \) & Number of allocated small models. \\
        \( N_{\text{large}} \) & Number of allocated large models. \\
        \( K_p, K_i, K_d \) & PID tuning parameters for dynamic resource allocation. \\
        \bottomrule
    \end{tabular}
\end{table}

\subsection{Generation using Cached Image with Reduced De-Noising Steps} \label{sec:generation_using_cache}

Upon receiving a request, the request scheduler uses the prompt embedding to check for a sufficiently similar image in the cache. If a match is found, \THISWORK\ retrieves the image, adds noise to it by a pre-determined amount, and finally uses a diffusion model to de-noise the image for the given prompt. The key idea is to augment with sufficient noise to generate a new image, enabling the system to skip de-noising steps. 
% The following details describe this mechanism further.

% Given a query embedding, the system retrieves the most relevant cached image based on cosine similarity.
% The image embedding is extracted using a pretrained CLIP image encoder, ensuring retrieval considers both semantic and structural similarity.
% If the similarity score surpasses a predefined threshold, the system classifies the request as a cache hit.
Given a query embedding \( q \), the system retrieves the most relevant cached image by computing the cosine similarity between the query embedding and cached image embeddings (following \S\ref{sec:how_to_retrieve}).
The similarity score is defined as:

\begin{equation}
    S(q, I^*) = \frac{q \cdot e_{I^*}}{\|q\| \|e_{I^*}\|}
\end{equation}

where \( e_{I^*} \) is the embedding of the cached image \( I^* \), extracted using a pretrained CLIP image encoder. Retrieval is performed only if the similarity score \( S(q, I^*) \) exceeds a predefined threshold \( \tau \), which is treated as a hyperparameter controlling the trade-off between retrieval precision and recall.
Rather than using the retrieved image directly, the system reintroduces noise using the same scaling method as image-to-image diffusion, enabling the image to re-enter the de-noising process at an intermediate timestep. This approach allows for refinement of the retrieved image, ensuring better alignment with the request while preserving computational efficiency.
Given a target timestep \( t_k \), the noisy image \( \tilde{I} \) is generated using the diffusion model’s noise schedule:
% \vspace{-0.125cm}
\begin{equation}
    \tilde{I} = \sigma_{t_k} \cdot \epsilon + (1 - \sigma_{t_k}) \cdot I^*,
\end{equation}
where \( \sigma_{t_k} \) is the noise scaling factor retrieved from the diffusion model’s noise schedule \( \text{sigmas}[t_k] \), \( \epsilon \sim \mathcal{N}(0, I) \) is Gaussian noise sampled from a standard normal distribution, and \( I^* \) is the retrieved image.

The system then runs the diffusion model for the remaining \( T - k \) steps, skipping the initial \( k \) de-noising iterations. This step refines the retrieved image to match the new request.
Since the retrieved image shares high-level features with the target, running only the later de-noising steps allows controlled adjustments to colors, textures, and details without regenerating the image from scratch.
By avoiding redundant computation, this method significantly reduces inference cost while preserving high image fidelity.

The choice of \( k \) is determined dynamically based on the similarity between the retrieved image and the new request (\S\ref{sec:cache_retrieval}). A higher similarity score indicates that the retrieved image closely matches the new prompt, requiring fewer modifications to align with the desired output. In such cases, a larger \( k \) is selected to skip more de-noising steps and reduce latency. Conversely, for lower similarity, a smaller \( k \) is used to allow more iterative refinement. 

% By reusing retrieved images and skipping the first \( k \) de-noising steps, our system eliminates a significant portion of the computational overhead.
With \( H_{\text{cache}} \) to be the cache hit rate with \( C_{\text{gen}} \) denoting the total compute cost for generating a new image from scratch using the large model. Each request requires a total of \( T \) de-noising steps with cache hits only requiring \( T - k \) steps. The computation saved per cache-hit request, weighted by the distribution of different diffusion steps \( K \), is:

\begin{equation}
    C_{\text{saved}} = H_{\text{cache}} \sum_{k=0}^{T} \frac{k}{T} C_{\text{gen}} P(K = k).
\end{equation}

Instead of running the large model for refinement, our system can offload cache-hit refinements to a smaller model, which performs the remaining \( T - k \) steps at a significantly lower cost. Let \( C_{\text{small}} \) denote the compute cost of the small model per step relative to the large model. The total compute savings, considering the distribution of \( K \), is:

\begin{equation}
    \resizebox{0.95\hsize}{!}{$
    C_{\text{total\_saved}} = H_{\text{cache}} \sum_{k=0}^{T} P(K = k) \left( \frac{k}{T} C_{\text{gen}} + \frac{T - k}{T} (C_{\text{gen}} - C_{\text{small}}) \right).
    $}
\end{equation}

Since the small model also performs only the remaining \( T - k \) steps, its impact on total compute savings depends on how much cheaper each step is relative to the large model. The larger the gap between \( C_{\text{gen}} \) and \( C_{\text{small}} \), the greater the additional efficiency gained by refinement offloading.

\subsection{Cache Retrieval and \( k \) Selection} \label{sec:cache_retrieval}
% To minimize redundant computation, our system employs a cache-aware retrieval mechanism that efficiently selects previously generated images for reuse.
% When a new request arrives, its prompt embedding is computed using a pre-trained CLIP model and compared against stored image embeddings.
% For cache hit, the system retrieves the corresponding cached image and determines an appropriate refinement parameter \( k \), which represents the amount of noise to add and number of de-noising steps that can be skipped for generation.

Since our system caches full images rather than latents, we can generate intermediate states for any \( k \) by applying controlled noise to the cached image.
To balance computational efficiency and generation diversity, we restrict \( k \) to choose from a discrete set of N values (where \( N = 6 \) in our case), \( \mathcal{K} = \{5, 10, 15, 20, 25, 30\} \) .
This selection ensures efficient refinement while covering a sufficient range of de-noising steps to maintain high-quality generation.
Given that a full generation requires (\( T = 50 \)) de-noising steps, we cap \( k \) at 30 to prevent excessive similarity between refined outputs while still achieving significant computational savings.
% Given this set of possible \( k \)-values, the retrieval algorithm selects the highest feasible \( k \) based on text-to-image similarity, allowing the system to minimize computational cost while maintaining output quality.

To ensure that cached image refinements maintain a sufficiently high level of quality, \THISWORK\ follows a quality-constrained retrieval policy.
Specifically, we enforce that the quality of an image generated from a retrieved cached image using a small model must be at least \( \alpha \) times the quality of an image generated from scratch using a large model, where \( \alpha \leq 1 \) is the quality degradation factor controlling the allowable trade-off between efficiency and output fidelity:
\begin{equation}
    Q_{\text{cache-hit}}(k) \geq \alpha \cdot Q_{\text{full-gen}}.
\end{equation}

We conduct an empirical evaluation to determine appropriate cache hit thresholds \( \tau \) and \( k \) selection values.
We generate 10000 images from DiffusionDB~\cite{wang2022diffusiondb} using a large model and then performed refinements starting at different \( k \) steps (\( k \in \{5, 10, 15, 20, 25, 30\} \)), selecting the most semantically aligned cached image from a cache of 100000 stored images.
We set a strict quality constraint of \( \alpha = 0.95 \).
While this analysis is inspired by prior work~\cite{nirvama}, the key difference is that we rely on text-to-image similarity for retrieval based on \S\ref{sec:how_to_retrieve} instead of earlier text-to-text similarity attempt.

\begin{figure}[t]
    \centering
    \begin{subfigure}[t]{0.65\linewidth}
        \centering
        \includegraphics[width=\linewidth]{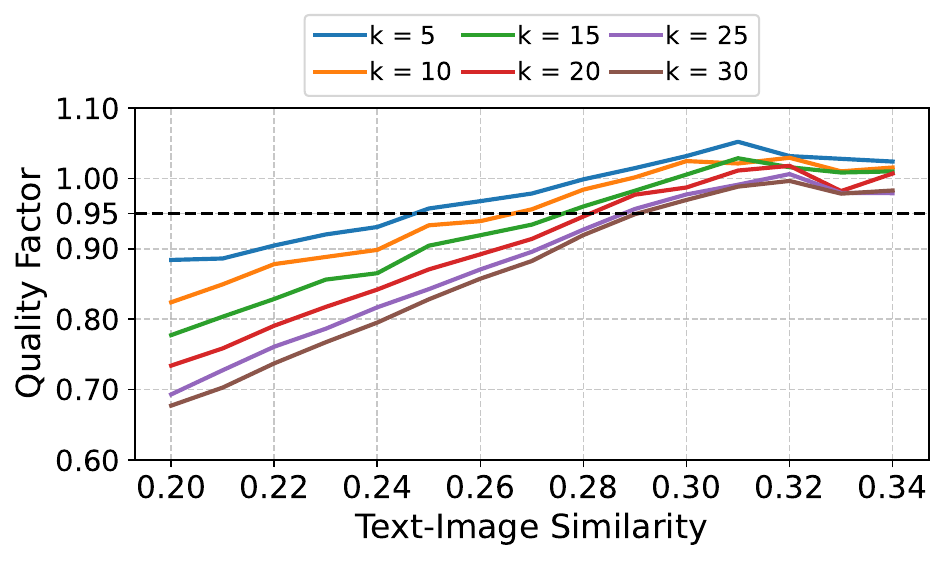}
        \vspace{-0.6cm}
        \caption{
        % Final image quality score vs. text-image similarity across different $k$.
        }
        \label{fig:cache_hit_threshold}
    \end{subfigure}
    \hfill
    \begin{subfigure}[t]{0.34\linewidth}
        \centering
        % \vspace{0pt} % This ensures top alignment
        \includegraphics[width=\linewidth]{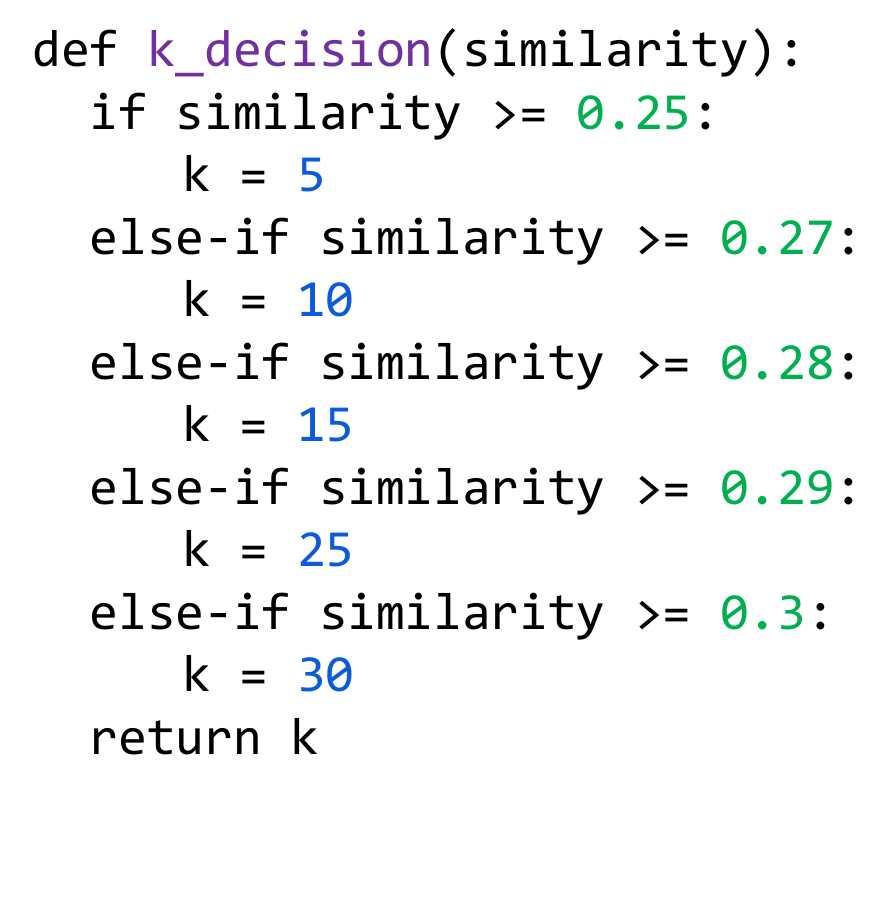}
        \vspace{-0.6cm}
        \caption{
        % $k$-selection based on text-to-image similarity.
        }
        \label{fig:k_logic}
    \end{subfigure}
    \vspace{-0.5cm}
    \caption{(a) Relationship between final image quality and text-image similarity across different $k$, (b) logic for determining $k$ based on text-to-image similarity.}
    \vspace{-0.5cm}
    \label{fig:combined_fig}
\end{figure}

We present a detailed analysis of these results in Fig.~\ref{fig:cache_hit_threshold}, illustrating the relationship between the similarity score between the query and retrieved image, the selected \( k \), and the final image quality.
For a fixed \( k \), as the text-image similarity increases, the final image quality also improves. Furthermore, for the same text-image similarity, a smaller \( k \) involves more refinement to the retrieved image, resulting in higher final image quality but lower computational savings. 
A quality factor greater than 1 is observed when the retrieved images closely align with the new prompt (\textit{i.e.,} exhibits high text-to-image similarity).
In such cases, even partial refinement using a smaller model can further enhances image quality.
The figure also marks the quality degradation parameter \( \alpha \), and for each \( k \), we determine the lowest possible text-image similarity that ensures the final image quality remains above \( \alpha \), using this value as the cache hit threshold for that specific \( k \).
Fig.~\ref{fig:k_logic} presents the decision process for determining \( k \) based on similarity.
This ensures that \THISWORK\ maximizes computational savings while maintaining the required image quality.

% \begin{figure}[t]
%     \centering
%     \includegraphics[width=0.8\linewidth]{Figures/k_logic.pdf}
%     \caption{K determination logic}
%     \label{fig:k_logic}
% \end{figure}

\textbf{Cache Threshold Selection.}
\THISWORK\ utilizes text-to-image similarity (CLIPScore) to achieve higher cache utilization while maintaining a stringent quality factor (\( \alpha \) $\geq0.95$) compared to \textsc{Nirvana}'s text-to-text similarity-based method for (\( \alpha \) $\geq0.90$).
Although \textsc{Nirvana} applies high thresholds (0.65-0.95) on text-to-text similarity, this measure does not directly account for the perceptual fidelity between the input prompt and generated images. In contrast, CLIP scores explicitly capture semantic alignment between text and images, providing a more accurate representation of the final visual output's relevance to the user's query. Consequently, even with lower numerical thresholds (from 0.25 to 0.3), \THISWORK\ achieves better cache utilization by targeting semantically similar images, better matching the user's intent.
% We also evaluate the effectiveness of the proposed threshold heuristic (Fig.~\ref{fig:k_logic}) by randomly sampling a new set of 1000 prompts that are not seen during the training for this heuristic and hit in the cache.
% For every prompt, we determine the number of denoising steps using our heuristic, generated the image, and measured its quality using the CLIP score.
% On average, our heuristic‐guided images scored 28.50, versus 28.59 for the full large model pipeline, \textit{i.e.,} 99.7\% of the baseline quality.
% This not only exceeds our 95\% quality retention target but also demonstrates that we can substantially reduce denoising iterations (and thus inference cost) with negligible loss in perceptual fidelity.
We evaluate the effectiveness of the proposed threshold heuristic (Fig.~\ref{fig:k_logic}) by randomly sampling a new set of 1,000 prompts, distinct from those used during training the heuristic, that result in cache hits.
For each prompt, we use our heuristic to determine the number of denoising steps, generate the corresponding image, and assess its quality using the CLIP score.
On average, the heuristic-guided images achieve a CLIP score of 28.50, compared to 28.59 from the full large-model pipeline, corresponding to 99.7\% of the baseline quality.
This not only surpasses our target of 95\% quality retention but also demonstrates that our method can significantly reduce the number of denoising iterations, and hence inference cost—with minimal loss in perceptual fidelity.

\textbf{Performance of Cache Retrieval.}
\THISWORK\ performs text-to-image similarity computations on the GPU with minimal overhead.
First, image embeddings require very little memory: storing 100000 image embeddings consumes just 0.29 GB.  
Second, cosine similarity computation is highly optimized for GPUs, as it involves element-wise normalization and matrix multiplication, both of which are efficient operations.  
The latency of cache retrieval, therefore, is negligible, taking only 0.05 seconds for 100000 cached images, whereas the de-noising process takes over 10 seconds.  
This ensures that similarity checks remain lightweight and do not become a bottleneck in cache retrieval.

\subsection{Resource Management using Global Monitor} \label{sec:global_monitor}
A large-scale serving framework must leverage multiple GPUs to efficiently handle high request volumes. Since diffusion models are computationally intensive, a single model running on a single GPU would quickly become a bottleneck under heavy workloads. To achieve high throughput, a serving system distributes requests across multiple GPUs, allowing concurrent execution of multiple inference tasks.
MoDM introduces a mixture-of-models design to balance latency and image quality. This necessitates a resource management system that efficiently allocates GPU workers between large and small models to balance request throughput, latency, and quality.
In specific we present two operational modes.

\begin{itemize}[leftmargin=*]
    \item \textbf{Quality-Optimized Mode:}
    \THISWORK\ aims to meet the request rate while maintaining the highest image quality.
    \item \textbf{Throughput-Optimized Mode}:
    \THISWORK\ maximizes throughput, ensuring the highest number of processed requests while still maintaining acceptable image quality.
\end{itemize}

\subsubsection{Quality-Optimized Mode}

In this mode, the resource management system dynamically allocates available GPU workers between large and small models to (1) meet the request rate requirement and (2) maintain image quality as high as possible.
The system determines the optimal number of large \( N_{\text{large}} \) and small  \( N_{\text{small}} \) models based on real-time metrics such as the cache hit rate, the distribution of cache-hit refinement steps \( k \), and the request rate \( R \).
Each GPU (a worker) can only host one model at a time, imposing

\begin{equation} \label{constraint_1}
    N_{\text{large}} + N_{\text{small}} \leq N.
\end{equation}
%Here, \( N \) represents the total number of workers.

The system must also satisfy the following constraints to ensure compliance with the latency objectives.

\textbf{Cache Miss Throughput Constraint.} 
Since cache-miss requests require full image generation and can only be processed by large models, the aggregate throughput of all large models must be at least the workload required to process cache miss requests at a given request rate.
\begin{equation} \label{constraint_2}
    N_{\text{large}} P_{\text{large}} \geq W_{\text{miss}} = (1 - H_{\text{cache}}) R.
\end{equation}

\textbf{Cache Hit Throughput Constraint.}
Cache hit requests are processed using either a large or a small model; the amount of work depends on the selected refinement step \( k \).
The effective workload for cache hit requests depends on the portion of requests assigned to different \( k \)-values:

\begin{equation} \label{constraint_3}
    W_{\text{hit}} = H_{\text{cache}} R \sum_{k \in \mathcal{K}} P(K = k) \left(1 - \frac{k}{T} \right).
\end{equation}

To ensure the system meets this workload demand, the remaining throughput from large  (after serving cache-miss requests) and the  small models must \textit{together} be at least the cache-hit workload.
\begin{equation} \label{constraint_4}
    \left( N_{\text{large}} P_{\text{large}} - W_{\text{miss}} \right) + N_{\text{small}} P_{\text{small}} \geq W_{\text{hit}}.
\end{equation}
% \begin{equation} \label{constraint_4}
%      N_{\text{small}} P_{\text{small}} \geq W_{\text{hit}}.
% \end{equation}

\textbf{Optimization Objective.} To maintain the highest possible image quality, we maximize the number of large models \( N_{\text{large}} \), subject to the constraints in Eqs. (5)–(8):

\begin{equation} \label{opt_obj}
    \max_{N_{\text{large}}, N_{\text{small}}} N_{\text{large}}, \quad 
    \textit{s.t.} \quad \text{Constraints~(\ref{constraint_1})–(\ref{constraint_4})}.
\end{equation}

\subsubsection{Throughput-Optimized Mode}

In this mode, all cache miss requests are processed using a large model, while all cache hit requests are processed using a small model. This strategy minimizes the total computational workload by leveraging the efficiency of small models for all refinement tasks.  
To achieve the highest possible throughput, the system balances the allocation of large and small models based on the ratio of cache-hit and cache-miss workloads. Since all cache-hit requests are processed by a small model, the cache-hit workload should be adjusted to account for the difference in throughput between large and small models. The weighted cache-hit workload is computed as:

\begin{equation}
    W_{\text{hit}}^{\text{weighted}} = \frac{W_{\text{hit}}}{P_{\text{small}} / P_{\text{large}}}.
\end{equation}
Here, \( W_{\text{hit}} \) is calculated in Eq.~(\ref{constraint_3}). 

The number of large models needed based on workload:

\begin{equation}
    N_{\text{large}} = \frac{W_{\text{miss}}}{W_{\text{hit}}^{\text{weighted}} + W_{\text{miss}}} \times N.
\end{equation}
where \( N \) represents the total number of available workers and \( W_{\text{miss}} \) is calculated in Eq.~(\ref{constraint_2}).

% To achieve our design objective, we design the global monitor algorithm that implements a dynamic resource allocation strategy optimizing the distribution of GPUs between large and small diffusion models.
% The algorithm starts by analyzing the request pattern from the previous recording period to compute two key workloads: non-hit workload, consisting of cache-miss requests that must be processed by large models, and hit workload, which includes cache-hit requests that can be handled by either model type with varying refinement steps.
% First, it determines the minimum number of large models required to satisfy the cache-miss throughput constraint (Eq.~\ref{constraint_2}), ensuring sufficient capacity for full image generation.
% Then, through an iterative process, it attempts to maximize the number of large models (as per the optimization objective in Eq.~\ref{opt_obj}) while verifying that the combined throughput from any remaining large model capacity plus all small models satisfies the cache-hit workload constraint (Eq.~\ref{constraint_4}).
% The algorithm (shown in Algorithm~\ref{alg:global_monitor}) incorporates a PID controller with carefully tuned parameters (Kp=0.6, Ki=0.05, Kd=0.05) to smooth transitions between configurations, preventing oscillations in resource allocation.
% This approach ensures high-quality image generation while maintaining system responsiveness under varying request patterns and workload distributions, effectively balancing the trade-off between generation quality and processing speed.

To achieve our design objective, we develop the Global Monitor algorithm (shown in Algorithm~\ref{alg:global_monitor}). The algorithm begins by analyzing request patterns from the previous recording period to compute two key workloads: the cache miss workload, which consists of requests that bypass the cache and must be fully processed by large models, and the cache hit workload, which includes requests that retrieve an image from the cache and require additional refinement. 

For quality-optimized mode, the algorithm first computes the minimum number of large models required to meet the cache miss throughput constraint (Eq.~\ref{constraint_2}), ensuring sufficient capacity for full image generation. It then iteratively increases the number of large models while verifying that the combined throughput of the remaining large and small models remains sufficient to handle the cache hit workload (Eq.~\ref{constraint_4}). This process continues until the cache-hit workload constraint is violated, aligning with the optimization objective in Eq.~\ref{opt_obj}.

For throughput-optimized operation, the system first computes the weighted cache hit workload to account for differences in throughput between large and small models. It then balances the allocation of large and small models proportionally to workload demands, ensuring an optimal distribution. This approach enables efficient scaling while preventing resource underutilization.

After obtaining the heuristic-based initial allocation for either mode, the PID controller acts as a stabilizer to maintain stability in resource allocation.
It continuously adjusts the number of large models based on real-time demand fluctuations, ensuring smooth adaptation to changing workload conditions.
With carefully tuned parameters (\( K_p = 0.6 \), \( K_i = 0.05 \), \( K_d = 0.05 \)), the controller refines the initial allocation by dampening rapid changes, preventing abrupt shifts or oscillations in GPU allocation.
This hybrid approach leverages the efficiency of heuristics for quick decision-making while utilizing PID to enforce gradual, stable adjustments, ensuring both responsiveness and robustness in dynamic serving environments.

\begin{algorithm}[t]
\scriptsize
\caption{Global Monitor for Dynamic Model Allocation}
\label{alg:global_monitor}
\begin{algorithmic}[1]
\Require Number of GPUs \( N \), PID parameters \( (K_p, K_i, K_d) \), large model throughput \( P_{\text{large}} \), small model throughput \( P_{\text{small}} \), total denoising steps \( T \), cache hit rates \( H_{\text{cache}} \), request rate \( R \), refinement step distribution \( k\_rates \)
\Ensure Dynamic allocation of large and small models to balance request load

\State \quad  Initialize \texttt{PIDController} with \( K_p, K_i, K_d \)
% \State \quad  Set \texttt{current\_num\_large} \( \gets N \)  \Comment{Initially allocate all GPUs to large models}
% \State Set \texttt{prev\_time} \( \gets \) \texttt{time.time()}

\State \quad \textcolor{blue}{\texttt{// Compute cache miss workload}}
\State \quad \texttt{miss\_workload} \( \gets (1 - H_{\text{cache}}) \times R \)

\State \quad \textcolor{blue}{\texttt{// Compute refinement workload factor}}
\State \quad \textbf{For each} \texttt{k, rate} \textbf{in} \texttt{k\_rates.items():}
\State \quad \quad \( F_{\text{refine}} \gets F_{\text{refine}} + \texttt{rate} \times \left(1 - \frac{k}{T} \right) \)

\State \quad \textcolor{blue}{\texttt{// Compute hit\_workload}}
\State \quad \texttt{hit\_workload} \( \gets H_{\text{cache}} \times R \times F_{\text{refine}} \)

\State \quad \textbf{If Quality-Optimized Mode:}
\State \quad \quad \textcolor{blue}{\texttt{// Compute minimum number of large models}}
\State \quad \quad \texttt{num\_large} \( \gets \lceil \texttt{miss\_workload} \,/\, P_{\text{large}} \rceil \)

\State \quad \quad \textcolor{blue}{\texttt{// Search for the maximum possible number of large models}}
\State \quad \quad \textbf{While} \texttt{num\_large} \( \leq \) \texttt{N} \textbf{do}  
\State \quad \quad \quad \texttt{available\_throughput} \( \gets \) \texttt{num\_large} \( \times \,\, P_{\text{large}} - \texttt{miss\_workload} \)
\State \quad \quad \quad \quad \quad \quad \quad \quad \quad \quad \quad \quad \quad  \quad\( + (\texttt{N} - \texttt{num\_large}) \times P_{\text{small}} \)

\State \quad \quad \quad \textbf{If} \texttt{available\_throughput} \( \geq \) \texttt{hit\_workload} \textbf{then}  
\State \quad \quad \quad \quad Increase \texttt{num\_large} by 1 and continue  
\State \quad \quad \quad \textbf{Else}  
\State \quad \quad \quad \quad Decrease \texttt{num\_large} by 1 and break  

\State \quad \textbf{Else If Throughput-Optimized Mode:}
\State \quad \quad \textcolor{blue}{\texttt{// Compute weighted cache-hit workload}}
\State \quad \quad \texttt{hit\_workload\_weighted} \( \gets \texttt{hit\_workload} \times \frac{P_{\text{large}}}{P_{\text{small}}} \)

\State \quad \quad \textcolor{blue}{\texttt{// Compute number of large models based on workload ratio}}
\State \quad \quad \texttt{num\_large} \( \gets \left( \frac{\texttt{miss\_workload} }{\texttt{hit\_workload\_weighted} + \texttt{miss\_workload}} \right) \times \texttt{N} \)

\State \quad \textcolor{blue}{\texttt{// Apply PID adjustment to \texttt{current\_num\_large}}}
\State \quad \( \Delta_{\text{large}} \gets \texttt{PIDController.compute(num\_large, current\_num\_large)} \)

\State \quad \texttt{current\_num\_large} \( \gets \texttt{current\_num\_large} + \Delta_{\text{large}} \)

\State \quad \textcolor{blue}{\texttt{// Compute final allocations}}
\State \quad \( N_{\text{large}} \gets \max(1, \min(\texttt{round(current\_num\_large)}, N)) \)

\State \quad \textbf{Return}
\State \quad \quad \texttt{model\_allocation} \( \gets \) \texttt{["large"]} \( \times N_{\text{large}} \) \( + \) \texttt{["small"]} \( \times (N - N_{\text{large}}) \)
\end{algorithmic}
\vspace{-0.1cm}
\end{algorithm}

% The global monitor dynamically allocates GPUs between different model variants to optimize throughput and system efficiency. It continuously tracks key metrics such as **request rate, cache hit rate, and the distribution of retrieved refinement steps (\( k \)-values)** to make informed scheduling decisions. A **PID (Proportional-Integral-Derivative) controller** is employed to adjust the number of GPUs assigned to each model variant in real time, ensuring **adaptive resource balancing** based on workload characteristics.

% When request traffic increases, the global monitor **allocates more GPUs to larger models** to handle cache-miss requests efficiently. Conversely, when a high proportion of cache-hit requests is observed, it **reallocates GPUs to smaller models** that specialize in refinement, optimizing computational efficiency. By dynamically adapting model allocation based on workload variations, the system maximizes **resource utilization while preventing GPU bottlenecks**.

\subsection{Cache Maintenance} \label{sec:cache_maintainance}
% Efficient cache management is critical for maintaining retrieval accuracy and system responsiveness. Since the number of cached images is limited by storage constraints, our system implements a **priority-based eviction strategy** that selectively removes less useful cached entries. When the cache reaches its maximum size, **the least valuable stored images are evicted** based on a combination of retrieval frequency and computational benefit.

% In our system, both cache-miss and cache-hit requests are added to the cache, enhancing the overall request hit rate, particularly when the maximum cache size is constrained.  
To maintain cache relevance, we explore trade-offs between different cache maintenance strategies. 
While prior work~\cite{nirvama} adopts a utility-based approach, we investigate whether a simpler sliding window-based strategy can be equally effective.  
In this sliding window approach, newly generated images update the cache while the oldest images are discarded, effectively following a FIFO management policy.  
Our findings indicate that this FIFO-based strategy performs well.  
Using the DiffusionDB~\cite{wang2022diffusiondb} production dataset, we analyze the temporal correlation between requests, measuring the time gap between a new request and its corresponding retrieved image.  
Fig.~\ref{fig:time_difference} (relocated to \S\ref{sec:timing_analysis} due to space limitation) presents this data, revealing that over 90\% of cache-hit requests retrieve images cached within four hours of the new request.  
This aligns with user behavior, as many users iteratively prompt diffusion models with similar text variations to refine generated content.  
Beyond its quantitative benefits, this policy also helps maintain content diversity in the cache.  
In contrast, a utility-based cache can lead to high reuse of a small subset of images, biasing future image generations.
Based on both qualitative and quantitative analysis, \THISWORK\ adopts a \textit{simple yet effective FIFO-based cache maintenance}.

As inference serving progresses in production, a key design decision is whether to cache only images generated by the large model or to include those generated by both the small and large models.  
To address this, we evaluate whether caching images produced by the small model maintains image generation quality.  
\S\ref{sec:cache_small} provides a detailed quantitative analysis, demonstrating that caching images from the small model does not degrade generation quality.  
Combined with the earlier observation that temporally proximate requests benefit caching, our findings indicate that caching all images—regardless of model size—enhances overall system performance without compromising quality.  
Based on this, \THISWORK\ \textit{caches all images} generated at runtime.

\subsection{Model-Agnostic Caching for Flexible Serving} \label{sec:model_interoperability}
Prior works~\cite{nirvama,deepCache,layerCaching,patchedServe} improve diffusion model inference using caching but rely on internal, model-specific features (\textit{e.g.,} latent representations, activations, image patches), restricting them to a single model for serving.
In contrast, \THISWORK\ caches final images based on \S\ref{sec:what_to_cache}, a universal representation compatible across multiple model families.  
This allows \THISWORK\ to serve requests with different diffusion models, dynamically balancing performance and quality to adapt to workload demands, SLOs, and image quality constraints.

% For example, cached images may originate from a large model (\textit{e.g.,} Stable Diffusion-3.5), while de-noising can be performed by reintroducing noise to a cached image and de-noising using a separate model family (\textit{e.g.,} SANA-1.6B).  
% By supporting a mixture of diffusion models, \THISWORK\ provides flexibility in resource-constrained environments where request rates exceed what even a small model from one family (\textit{e.g.,} Stable Diffusion-XL) can handle.
% Our evaluation demonstrates a concrete use case where switching model families prevents Service Level Objective (SLO) violations under high request loads with a fixed number of GPU workers.

\section{Evaluation Methodology} \label{section:methodology}

\textbf{\underline{Implementation and Hardware.}}
We implement \THISWORK, our inference-serving system, in Python using PyTorch~\cite{ansel2024pytorch}, with the request scheduler, global monitor, and each worker running in a separate process.
Node communication is handled by PyTorch RPC~\cite{damania2023pytorch}, enabling efficient distributed deep learning. 
\THISWORK\ is deployed on two hardware configurations: a single server with four NVIDIA A40 GPUs (48GB memory) and a 16-node cluster, each node equipped with four AMD MI210 GPUs (64GB memory).

% We implement \THISWORK, our inference-serving system, in Python using PyTorch~\cite{ansel2024pytorch}, with a request scheduler, global monitor, and each worker managed by a single process. 
% Communication among nodes is handled using PyTorch RPC~\cite{damania2023pytorch}, a high-performance solution for distributed deep learning.
% We deploy \THISWORK\ on two different hardware configurations.
% The first setup consists of a single server with four NVIDIA A40 GPUs with 48GB on-device memory.
% The second configuration is a 16-node cluster, each node has four AMD MI210 GPUs with 64GB on-device memory.

% The first setup consists of a single server equipped with an Intel(R) Xeon(R) Platinum 8380 CPU and four NVIDIA A40 GPUs with 48GB VRAM. The second configuration is a 16-node cluster, where each node features two AMD EPYC 7V13 64-core CPUs and four AMD MI210 GPUs with 64GB VRAM.

\noindent\textbf{\underline{Models and Workloads.}}
We evaluate \THISWORK\ using four models to demonstrate cross-model compatibility.
The large models used in our study includes Stable Diffusion-3.5-Large (SD3.5L) with 8B parameters~\cite{sd3} and FLUX.1-dev (FLUX) with 12B parameters~\cite{blackforestlabs_flux1}.
The small models include Stable Diffusion XL (SDXL) with 3B parameters~\cite{sdxl} and SANA with 1.6B parameters~\cite{xie2024sana}.
SD3.5L and FLUX are both flow-based models trained using the Flow Matching framework.
We also include Stable Diffusion-3.5-Large-Turbo (SD3.5L-Turbo), a distilled variant of the SD3.5L model optimized to generate high-quality images in significantly fewer steps, to compare \THISWORK\ against a distilled baseline.
All models except SDXL run in BF16 precision, while SDXL uses FP16, following the developers' default recommendations.
All models use $T=50$ denoising steps and generate 1024$\times$1024 images, except SD 3.5-Turbo, which uses 10 steps.
For evaluation, we use DiffusionDB~\cite{wang2022diffusiondb}, a real-world production dataset with 2 million images, and MJHQ~\cite{mjhq_dataset}, a high-quality dataset of 30k MidJourney-generated images. 

\noindent
\textbf{\underline{Modeling of Request Arrivals.}}
% The similarity in requests stem from two reasons: (1) similar request from the same user, and (2) similar requests from different users.
% We refer the reader to NIRVANA \cite{nirvama} for a detailed study of request similarities that shows a nontrivial median cross-user similarity of $\approx$ 0.6, indicating abundant opportunities for semantic caching.
We model the request arrival process as a \textit{homogeneous Poisson process} with varying rates.
We replay the trace of user-submitted prompts from the DiffusionDB dataset in their original arrival order (and in trace order for MJHQ, which lacks timestamp information) to emulate a production environment consistent with the portal deployment of a Stable Diffusion–based model.
The image cache operates with the proposed FIFO-based cache management policy that shows a high degree of temporal locality as detailed in Appendix~\ref{sec:timing_analysis}.

\noindent \textbf{\underline{System Performance.}}
We use the following metrics.

\textbf{\textit{Maximum Throughput:}} We compare the highest throughput achieved by different baselines assuming that there enough number of requests to keep the system busy.

\textbf{\textit{Service Level Objective (SLO) Compliance}}: We analyze the system's ability to meet predefined latency thresholds under varying request rates. Specifically, we evaluate compliance with two SLO requirements: latency within \textbf{2$\times$} and \textbf{4$\times$} that of the large model (Stable Diffusion-3.5-Large) inference.

\textbf{\textit{Tail Latency}}: We measure the \textbf{99th percentile latencies} to capture the worst-case response times and ensure system stability under load.

\textbf{\textit{Maximum Load}}: We determine the highest sustainable request rate the system can handle while maintaining acceptable latency and quality of service.

\noindent \textbf{\underline{Image Quality.}}
Four metrics are used for quality.

\textbf{\textit{CLIPScore}}~\cite{clipScore}: Measures the alignment between the generated image and its corresponding prompt, providing an assessment of semantic accuracy.

\textbf{\textit{FID (Fréchet Inception Distance) Score}}~\cite{FID}: Quantifies the similarity between generated and groundtruth images by comparing their feature distributions.

\textbf{\textit{IS (Inception Score)}}~\cite{szegedy2016rethinking}: Evaluates the quality and diversity of generated images by analyzing the confidence and variety of class predictions from an Inception network.

\textbf{\textit{PickScore}}~\cite{kirstain2023pick}: Estimates human preference alignment by scoring how likely a generated image would be selected over others, trained on pairwise human comparisons.

\noindent \textbf{\underline{Baselines.}}
We compare against the following baselines.

\textbf{\textit{Vanilla System}}: An inference-serving system where every request is fully processed by the large model (\textit{i.e.,} SD3.5L or FLUX) without leveraging cached images or retrieval mechanisms. This represents a traditional approach to serving diffusion models.

\textbf{\textit{\textsc{Nirvana}}}~\cite{nirvama}: A caching-based diffusion inference system that improves efficiency by reusing previously generated images to reduce redundant computation.

\textbf{\textit{\textsc{Pinecone}}}~\cite{pinecone2023caching}: Retrieves and serves an image based on the most similar prompt using CLIP text embedding similarity without refinement; if no suitable match is found, it generates the image from scratch.
% \textsc{Nirvana} employs learned cache policies and retrieval mechanisms to optimize image reuse while maintaining quality.

% Additionally, we compare the image quality of \THISWORK\ against both baselines as well as images generated exclusively using the small model. This evaluation helps assess the trade-offs between efficiency and image fidelity.

\begin{figure}[t] 
    \centering 
    \includegraphics[width=0.48\textwidth, trim={0cm 0cm 0cm 1.6cm}, clip]{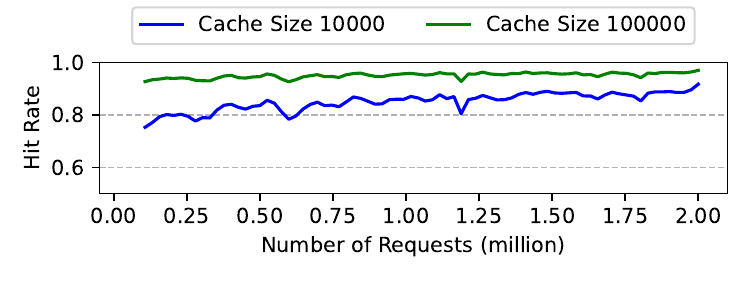}
    \vspace{-1cm}
    \caption{Hit rate of \THISWORK\ for DiffusionDB.}
    \vspace{-0.5cm}
    \label{fig:hit-rate-overall} 
\end{figure}

\noindent \textbf{\underline{Experimental Setup.}}
% We conduct experiments to evaluate tail latency, SLO compliance, throughput, and image quality.
Fig.~\ref{fig:hit-rate-overall} shows the overall hit rate across all 2 million requests from DiffusionDB with two different cache sizes.
The hit rate remains consistent across both settings, indicating that conducting experiments on a subset of the dataset provides a generalizable result.

\textbf{\textit{Throughput Experiments:}}  
We conduct experiments with 10000 requests from each dataset, using a cache of 10000 images generated by the large model. Unlike latency experiments, these tests focus solely on measuring maximum system throughput, ignoring timestamps. Additionally, we perform a scalability study w.r.t. GPU resources.
% For throughput evaluation, we conduct experiments using 10000 requests from each dataset, leveraging a cache of 10000 previously generated images by the large model. Unlike latency experiments, these tests do not consider request timestamps, focusing solely on measuring maximum system throughput. Additionally, we perform a scalability study to assess the system's performance as the serving cluster size increases.

\textbf{\textit{Latency and SLO Experiments:}}  
To evaluate tail latency and SLO compliance, we use  the DiffusionDB dataset, sorting them by arrival time.
We assign timestamps to requests using a Poisson distribution under different request rates to study the performance of the system under varying load.

\textbf{\textit{Image Quality Comparison:}}  
To assess image quality, we generate 10000 cached images per dataset in a warm-up phase, then serve another 10000 requests, retrieving cached images when possible. 
This experiment runs in throughput-optimized mode, representing the worst-case image quality scenario (\S\ref{sec:global_monitor}).
For FID, we generate four sets of 10000 images using the large model with same requests, different seeds, randomly selecting ground truth for comparison.

% To assess image quality, we first generate 10000 images using requests from each dataset as cached images in a warm-up phase. After caching these images, we serve another 10000 requests from each dataset, attempting to retrieve and reuse the cached images. For this experiment, we run the system in throughput-optimized mode, as it provides the worst-case image quality scenario for our system (\S\ref{sec:global_monitor}).
% For FID score, we generate four sets of 10000 images using the large model with the same requests but different seeds and randomly select images from these sets as ground truth for each request.

\section{Evaluation Results} \label{section:results}

We evaluate the performance of \THISWORK\ using a variety of metrics including throughput, adaptability to varying request rates, scalability with respect to GPU resources, SLO compliance, and the quality of image generation\footnote{Profiling results may vary across different software stack and configurations. The preliminary results presented here are specific to the evaluation framework used in this study for representative academic purpose.}.

% We evaluate MoDM in terms of throughput, scalability, tail latency, and SLO compliance across different request rates and hardware configurations.

\subsection{Throughput Evaluation} \label{sec:throughput_result}
\begin{figure}[t]
    \centering
    \includegraphics[width=\linewidth]{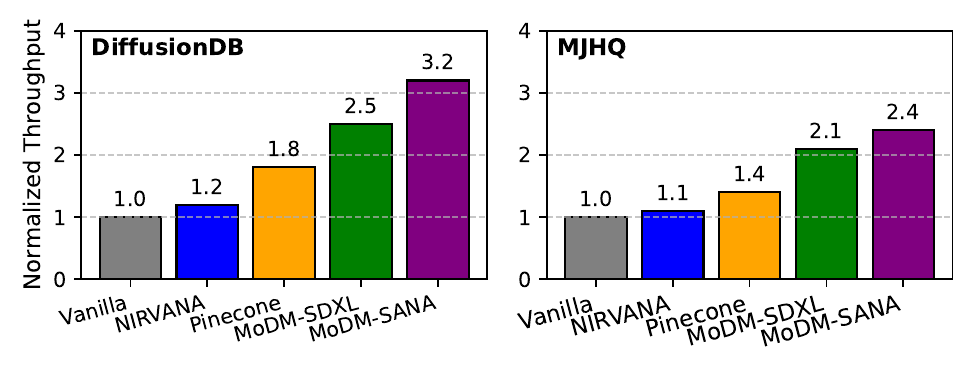}
    \vspace{-1cm}
    \caption{Throughput of different baselines normalized to Vanilla (Stable Diffusion-3.5-Large) on two datasets.}
    \label{fig:throughput}
\end{figure}

\begin{figure}[t]
    \centering
    \includegraphics[width=0.6\linewidth]{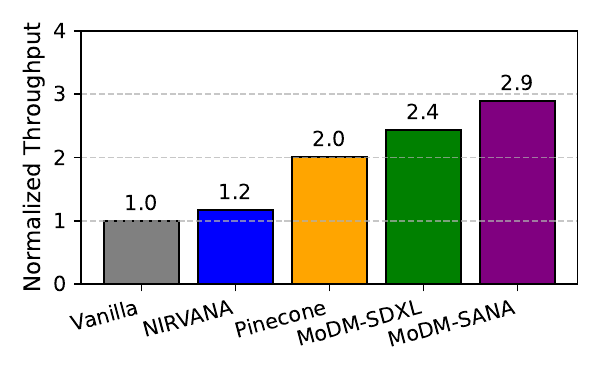}
    \vspace{-0.5cm}
    \caption{Throughput of different baselines normalized to Vanilla (FLUX) on DiffusionDB dataset. Both \textsc{Nirvana} and \THISWORK\ baselines use FLUX as a large model.}
    \label{fig:throughput_flux}
\end{figure}

Fig.~\ref{fig:throughput} presents the normalized throughput of \THISWORK\ and several baselines on the DiffusionDB and MJHQ datasets, using a cache size of 10,000 images. Throughput is normalized to the Vanilla baseline (Stable Diffusion-3.5-Large). On the real-world DiffusionDB workload, \THISWORK\ achieves a 3.2$\times$ improvement with SANA as the small model and 2.5$\times$ with SDXL. On MJHQ—a synthetic dataset with less temporal locality—the cache hit rate drops, resulting in reduced speedups (2.4$\times$ with SANA and 2.1$\times$ with SDXL). The performance gains primarily come from avoiding redundant denoising steps and utilizing smaller models on cache hits.
To evaluate generality across different large model baselines, Fig.~\ref{fig:throughput_flux} reports throughput on DiffusionDB normalized to FLUX.
\THISWORK\ continues to outperform all baselines, achieving up to 2.9$\times$ speedup with SANA and 2.4$\times$ with SDXL as small models.
This demonstrates the robustness of our approach under varying model configurations and workload characteristics.

\begin{figure}[t]
    \centering
    \includegraphics[width=\linewidth]{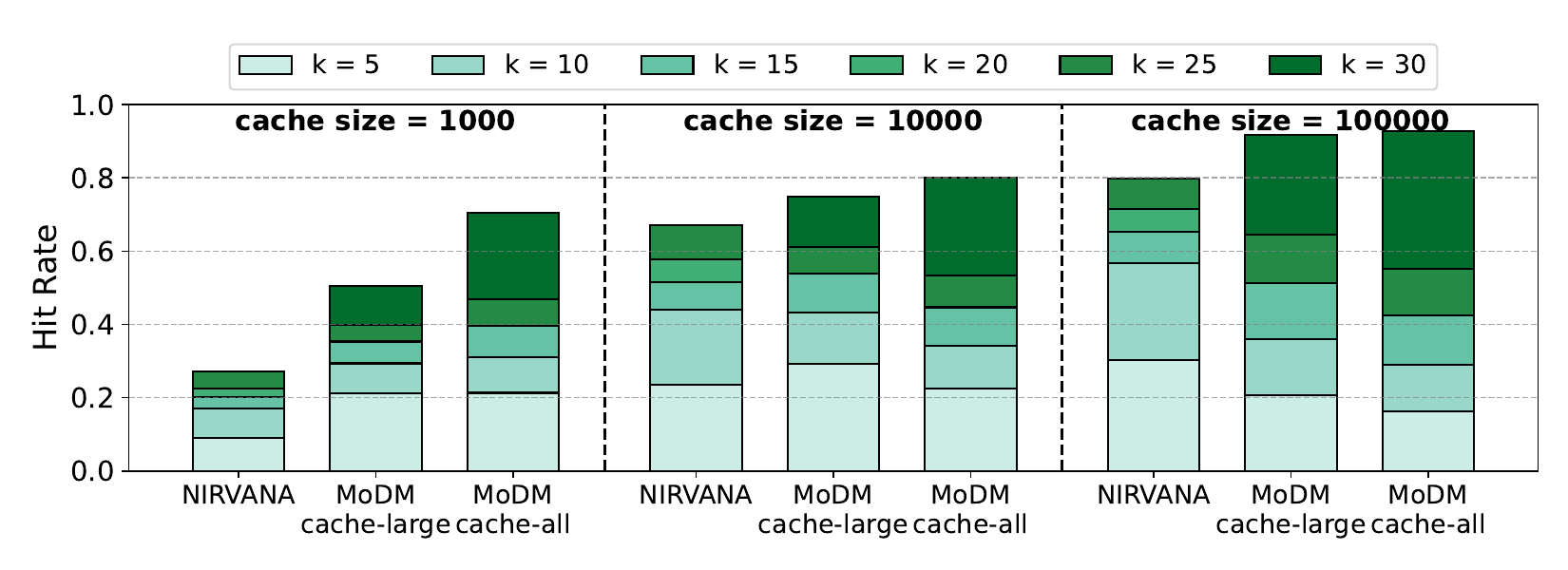}
    \vspace{-0.8cm}
    \caption{Comparison of cache hit rates for \textsc{Nirvana} and \THISWORK\ for the DiffusionDB dataset~\cite{wang2022diffusiondb}.}
    \label{fig:cache_policy_comparison}
\end{figure}

Fig.~\ref{fig:cache_policy_comparison} compares cache hit rates and skipped de-noising steps between \textsc{Nirvana} and \THISWORK\ using DiffusionDB (see \S\ref{sec:cache_hit_rates_mjhq} for the MJHQ dataset).
We evaluate \THISWORK\ under two configurations: (1) caching only images generated by the large model when a cache miss occurs (\THISWORK\ cache-large) and (2) caching both cache-miss images and refined cache-hit images from both small and large models (\THISWORK\ cache-all).
Cache hit rates are assessed across three cache sizes: 1000, 10000, and 100000.  

The results highlight four key insights.  
\textit{First, \THISWORK\ achieves higher cache hit rates than \textsc{Nirvana}.}  
This improvement stems from our sliding window-based cache maintenance policy, which better accommodates similar requests occurring in close temporal proximity (\S\ref{sec:cache_maintainance}), and our text-to-image similarity based retrieval policy, which enhances visual alignment with prompts (\S\ref{sec:how_to_retrieve}).  
\textit{Second, our text-to-image retrieval strategy increases the number of skipped de-noising steps.}  
A higher $k$ value (dark green portion) indicates more skipped steps, leading to substantial computational savings.

\textit{Third, caching images from both small and large models further improves cache hit rates.}  
This is expected, as caching all images better serves temporally adjacent requests, whereas caching only large-model-generated images results in missed opportunities.  
Additionally, as shown in \S\ref{sec:cache_small}, caching images generated by the small model does not degrade image quality, justifying this design choice from a quality standpoint.
\textit{Fourth, a cache size of 100000 is sufficient to achieve a high hit rate of 92.8\%.}  
This insight informs our design decision to balance cache size with hit rates while minimizing cache management and retrieval overhead.
Notably, this is much smaller than prior work~\cite{nirvama}.

\begin{figure}[t]
    \centering
    \includegraphics[width=\linewidth]{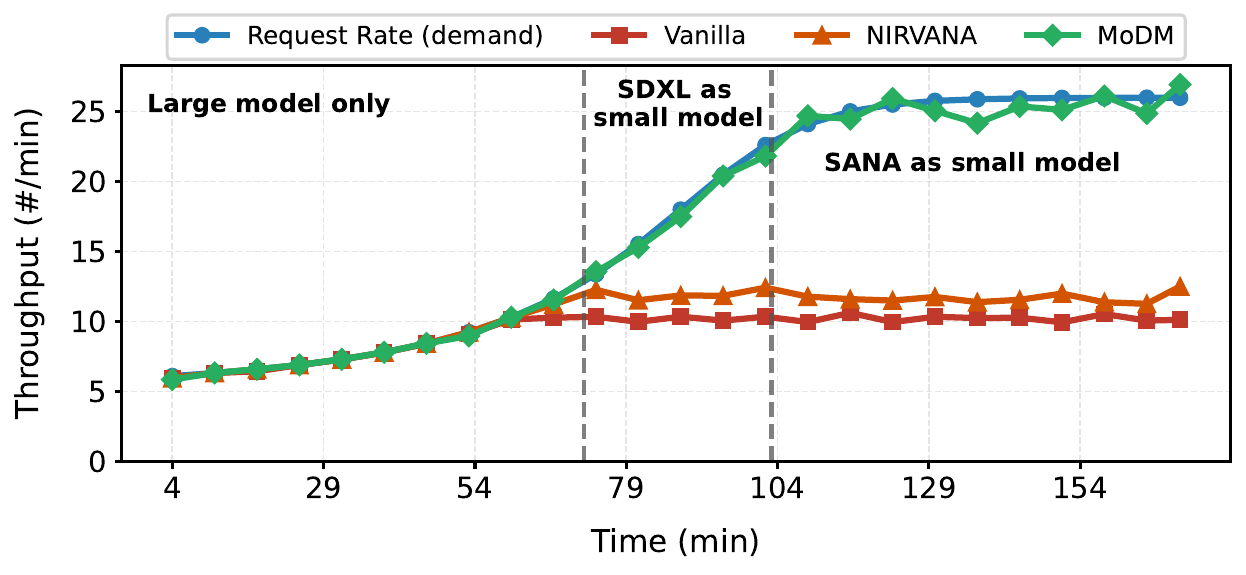}
    \vspace{-0.8cm}
    \caption{System throughput under increasing request rates.}
    \label{fig:throughput_climbing_rate}
\end{figure}

Fig.~\ref{fig:throughput_climbing_rate} illustrates how \THISWORK\ handles an increasing request rate, ranging from 6 to 26 requests per minute. 
This experiment was conducted using 16 MI210s.
The vanilla system reaches a maximum throughput of approximately 10 requests per minute, while \textsc{Nirvana} achieves a 20\% improvement.
However, as the request rate continues to rise, only \THISWORK\ is able to sustain the required throughput.
Between 12 and 22 requests per minute, \THISWORK\ uses SDXL as the small model for efficiency. Beyond 22 requests per minute, even SDXL cannot meet demand. 
To address this, \THISWORK\ dynamically switches the small model from SDXL to SANA, further increasing throughput.
We also evaluate these systems under fluctuating request rates, as shown in Fig.~\ref{fig:throughput_climbing_rate_f} in the Appendix.
\THISWORK\ adapts to changes in workload more effectively than baseline systems, maintaining high throughput throughout periods of variability.
\textit{\textbf{This adaptive model selection allows \THISWORK\ to maintain stable performance by dynamically balancing image quality and latency, effectively handling diverse use cases like varying SLO demands and quality requirements.
No prior state-of-the-art work offers such a broad and flexible trade-off between latency and quality.}}
% This adaptive model selection enables \THISWORK\ to maintain stable performance by dynamically adjusting image quality and latency that allows \THISWORK\ to better handle practical diverse use cases such as changing SLO demands and image quality requirements.
% No state-of-the-art prior work can dynamically adjust this trade-off to achieve a large design space between latency and quality.

% across varying loads, demonstrating its ability to efficiently scale under increasing traffic conditions.

\begin{figure}[t]
    \centering
    \includegraphics[width=\linewidth]{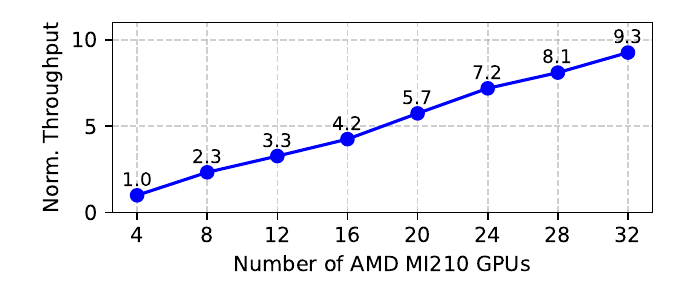}
    \vspace{-1cm}
    \caption{Scalability of \THISWORK\ with increasing \#GPUs.}
    \label{fig:scalability}
\end{figure}

To evaluate \THISWORK’s scalability, we analyze how throughput improves with respect to GPU resources. 
Fig.~\ref{fig:scalability} demonstrates super-linear scalability on MI210s, indicating that \THISWORK\ effectively utilizes available resources and remains adaptable to larger clusters without bottlenecks.
The observed super-linear scalability arises from the fact that, with more GPUs, requests are processed at a faster rate. As a result, within a given time period, a greater number of generated images are added to the cache, leading to a higher cache hit rate. This increased cache efficiency further boosts throughput, reinforcing the benefits of scaling multiple GPUs.

\subsection{SLO Compliance} \label{sec:slo_compliance}
\begin{figure}[t]
    \centering
    \includegraphics[width=\linewidth]{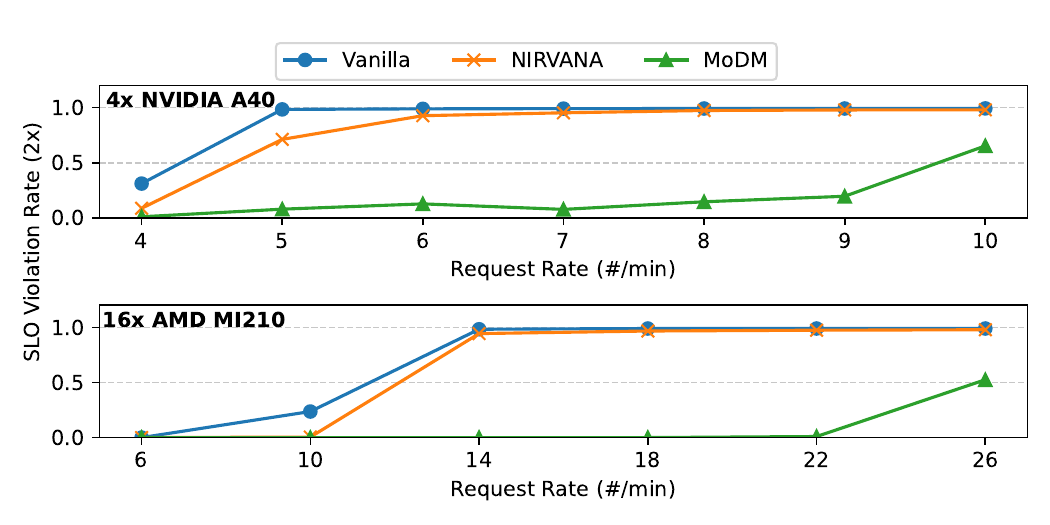}
    \vspace{-0.8cm}
    \caption{SLO violation rate (>2$\times$ inference latency of Stable Diffusion-3.5 Large) for different request rates.}
    \label{fig:SLO2x}
\end{figure}

\begin{figure}[t]
    \centering
    \includegraphics[width=\linewidth]{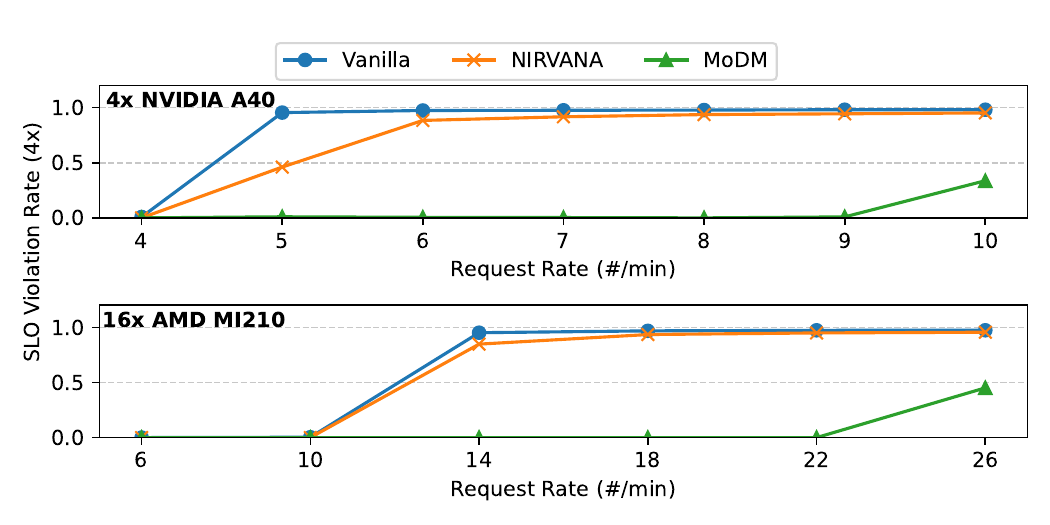}
   \vspace{-0.8cm}
    \caption{SLO violation rate (>4$\times$ inference latency of Stable Diffusion-3.5 Large) for different request rates.}
    \label{fig:SLO4x}
\end{figure}

Figs.~\ref{fig:SLO2x} and~\ref{fig:SLO4x} depict the SLO violation rates across different baselines.  
We evaluate SLO violations under two conditions: (1) when the generation latency exceeds 2$\times$ that of the large model (Stable Diffusion-3.5-Large), as shown in Fig.~\ref{fig:SLO2x}, and (2) when it surpasses 4$\times$ the large model’s latency, as shown in Fig.~\ref{fig:SLO4x}. 
Both figures compare SLO violations across varying request loads for two hardware configurations: 4 NVIDIA A40s and 16 AMD MI210s.  
\textit{The results demonstrate \THISWORK's ability to sustain significantly higher request loads without violating SLO, using the same hardware resources.}

At low request rates, all three systems meet SLO requirements with minimal violations.  
However, as the request rate exceeds 5 requests per minute on A40s and 14 requests per minute on MI210s, SLO violations become predominant in both the vanilla system and \textsc{Nirvana}.  
In contrast, \THISWORK\ maintains compliance for much higher loads, supporting up to 10 requests per minute on A40s and 22 requests per minute on MI210s under the 2$\times$ threshold, and up to 26 requests per minute on MI210s under the 4$\times$ threshold.
\THISWORK\ achieves this by leveraging a combination of large and small models. 
As shown in Fig.~\ref{fig:throughput_climbing_rate}, our system adaptively switches to a small diffusion model for inference under high request rates, significantly reducing computational overhead.
These results underscore \THISWORK’s superior efficiency in handling high request loads while minimizing SLO violations.
\S\ref{sec:tail_latency} in appendix expands on these results further showing the 99th percentile tail latency values for different baselines.

\subsection{Image Quality} \label{sec:image_quality}

% We evaluate the quality of generated images using CLIP Score and Fréchet Inception Distance (FID). The CLIP Score measures the semantic alignment between the generated image and its corresponding text prompt, with higher values indicating better alignment. The FID score assesses the realism and diversity of generated images by comparing their feature distributions with those of real images, where lower values indicate higher fidelity.

Table~\ref{tab:image_quality} compares the image generation quality of \THISWORK\ with various baselines on the DiffusionDB and MJHQ-30k datasets, using SD3.5L as the vanilla large model. Across both datasets, \THISWORK\ achieves CLIP, IS, and Pick scores comparable to the Vanilla baseline and \textsc{Nirvana}, demonstrating strong semantic alignment, perceptual diversity, and human preference alignment. Importantly, \THISWORK\ obtains substantially lower FID scores compared to standalone small or distilled models like SDXL, SD3.5L-Turbo, and SANA, indicating it preserves a high-quality distribution similar to the large model and avoids the occasional distortions and defects typical of small model outputs. In contrast, the \textsc{Pinecone} baseline shows noticeably lower CLIP scores, reflecting weaker image-text alignment and highlighting the limitations of retrieval-only methods without generative refinement—emphasizing the effectiveness of \THISWORK's refinement approach.

Separately, Table~\ref{tab:image_quality_flux} reports image quality on DiffusionDB using FLUX as the vanilla large model. Here, \THISWORK\ again balances quality and efficiency by achieving CLIP, IS, and Pick scores close to the strong FLUX baseline while improving upon the FID scores of standalone small or distilled models. This confirms that \THISWORK’s approach generalizes well across different large model backbones. Fig.~\ref{fig:appendix_images} in Appendix shows example images generated via various methods comparing with \THISWORK, clearly showcasing its high quality image generation ability.

% Table~\ref{tab:image_quality} presents the CLIP and FID scores for different baselines. The vanilla system achieves a CLIP Score of 28.55 and an FID of 6.29 on DiffusionDB. On MJHQ-30k, it attains a CLIP Score of 28.77 with an FID of 5.16. MoDM-SDXL achieves a CLIP Score of 28.70 on DiffusionDB and 28.77 on MJHQ-30k, closely matching the vanilla system in text-image alignment while having slightly higher FID values of 11.85 and 7.35. MoDM-SANA shows a CLIP Score of 28.01 on DiffusionDB and 28.82 on MJHQ-30k, maintaining reasonable alignment with prompts. However, its FID values are 16.96 and 10.51, reflecting a drop in realism that stems from the lower FID achieved by the SANA model itself.

% The standalone models SDXL and SANA exhibit notably higher FID values. SDXL has an FID of 16.29 on DiffusionDB and 12.67 on MJHQ-30k, while SANA reaches 19.96 and 16.31, respectively. These results indicate that MoDM achieves better feature distribution similarity to the vanilla system compared to standalone models, demonstrating that the retrieval and refinement process helps retain the characteristics of high-fidelity image generation.

\begin{table}[t]
    \centering
    \scriptsize
    \caption{Image quality on DiffusionDB and MJHQ-30k (Vanilla model: SD3.5L). Higher is better for CLIP, IS, and Pick; lower is better for FID.}
    \vspace{-0.3cm}
    \label{tab:image_quality}
    \renewcommand{\arraystretch}{1.2}
    \setlength{\tabcolsep}{3pt}
    \begin{tabular}{lcccccccc}
        \hline
        & \multicolumn{4}{c}{\textbf{DiffusionDB}} & \multicolumn{4}{c}{\textbf{MJHQ-30k}} \\
        \cline{2-9}
        \textbf{Baseline} 
        & \textbf{CLIP} $\uparrow$ & \textbf{FID} $\downarrow$ & \textbf{IS} $\uparrow$ & \textbf{Pick} $\uparrow$ 
        & \textbf{CLIP} $\uparrow$ & \textbf{FID} $\downarrow$ & \textbf{IS} $\uparrow$ & \textbf{Pick} $\uparrow$ \\
        \hline
        Vanilla (SD3.5L)    & 28.55 & 6.29 & 15.52 & 21.44 & 28.77 & 5.16 & 25.84 & 21.67 \\
        SDXL                & 29.30 & 16.29 & 16.90 & 21.45 & 29.66 & 12.67 & 25.82 & 21.55 \\
        SD3.5L-Turbo        & 27.23 & 14.63 & 15.38 & 21.45 & 27.84 & 10.68 & 23.70 & 21.59 \\
        \hline
        SANA                & 28.08 & 19.96 & 12.20 & 20.78 & 28.83 & 16.31 & 21.90 & 21.32 \\
        \textsc{Nirvana}    & 28.02 & 9.01  & 15.38 & 21.28 & 28.57 & 5.37 & 25.04 & 21.59 \\
        \textsc{Pinecone}    & 25.98 & 14.18  & 15.09 & 20.80 & 27.20 & 6.80 & 25.99 & 21.27 \\
        \hline
        \THISWORK-SDXL      & 28.70 & 11.85 & 15.27 & 21.00 & 28.79 & 6.87 & 25.46 & 21.33 \\
        % \THISWORK-Turbo      & 27.54 & 11.75 & 14.25 & 21.08 & 28.12 & 7.17 & 24.45 & 21.37 \\
        \THISWORK-SANA      & 28.01 & 16.96 & 12.67 & 20.79 & 28.82 & 9.96 & 22.25 & 21.28 \\
        \hline
    \end{tabular}
\end{table}

\begin{table}[t]
    \centering
    \scriptsize
    \caption{Image quality on DiffusionDB (Vanilla model: FLUX). Higher is better for CLIP, IS, and Pick; lower is better for FID.}

    \vspace{-0.3cm}
    \label{tab:image_quality_flux}
    \renewcommand{\arraystretch}{1.2}
    \setlength{\tabcolsep}{6pt}
    \begin{tabular}{lcccc}
        \hline
        \textbf{Baseline} 
        & \textbf{CLIP} $\uparrow$ 
        & \textbf{FID} $\downarrow$ 
        & \textbf{IS} $\uparrow$ 
        & \textbf{Pick} $\uparrow$ \\
        \hline
        Vanilla (FLUX)     & 26.82 & 6.02  & 16.69 & 21.29 \\
        SDXL                 & 29.30 & 17.60 & 16.90 & 21.45 \\
        SD3.5L-Turbo         & 27.23 & 15.11 & 15.38 & 21.45 \\
        \hline
        SANA                 & 28.08 & 24.37 & 12.20 & 20.78 \\
        \textsc{Nirvana}     & 26.01 & 9.07  & 15.44 & 21.06 \\
        \textsc{Pinecone}    & 24.37 & 19.41 & 16.08 & 20.63 \\
        \hline
        \THISWORK-SDXL       & 28.41 & 10.74 & 15.61 & 21.13 \\
        % \THISWORK-Turbo      & 27.54 & 11.75 & 14.25 & 21.08 \\
        \THISWORK-SANA       & 27.59 & 16.84 & 12.70 & 20.84 \\
        \hline
    \end{tabular}
\end{table}

%     \item \textbf{Performance of Cache-Hit Offloading:}
%     \begin{itemize}
%         \item \textbf{Throughput Improvements:}
%         \begin{itemize}
%             \item Increase in the number of requests served per minute by offloading cache-hit requests to small models.
%             \item Breakdown of GPU utilization across large and small models.
%         \end{itemize}
%         \item \textbf{Latency Reductions:}
%         \begin{itemize}
%             \item Reduced average response times for cache-hit requests.
%             \item Distribution of latencies (e.g., 90th percentile for cache hits vs. non-hits).
%         \end{itemize}
%         \item \textbf{Quality Preservation:}
%         \begin{itemize}
%             \item CLIP score analysis for cache-hit requests served by small models.
%             \item Comparison of quality trade-offs with increasing cache reliance.
%         \end{itemize}
%     \end{itemize}

%     \item \textbf{Sensitivity Analysis:}
%     \begin{itemize}
%         \item \textbf{Dynamic Model Allocation:}
%         \begin{itemize}
%             \item Analyze system performance under varying request rates.
%             \item Effectiveness of dynamic model loading policy in balancing throughput and image quality.
%         \end{itemize}
%     \end{itemize}

%     \item \textbf{SLO Compliance:}
%     \begin{itemize}
%         \item \textbf{Latency SLO:}
%         \begin{itemize}
%             \item Percentage of requests meeting latency objectives for both cache hits and non-hits.
%         \end{itemize}
%     \end{itemize}
% \end{itemize}

\subsection{Comparison on the Quality-Performance Trade-off Space} \label{sec:tradeoff}
\begin{figure}[h]
    \centering
    \includegraphics[width=\linewidth]{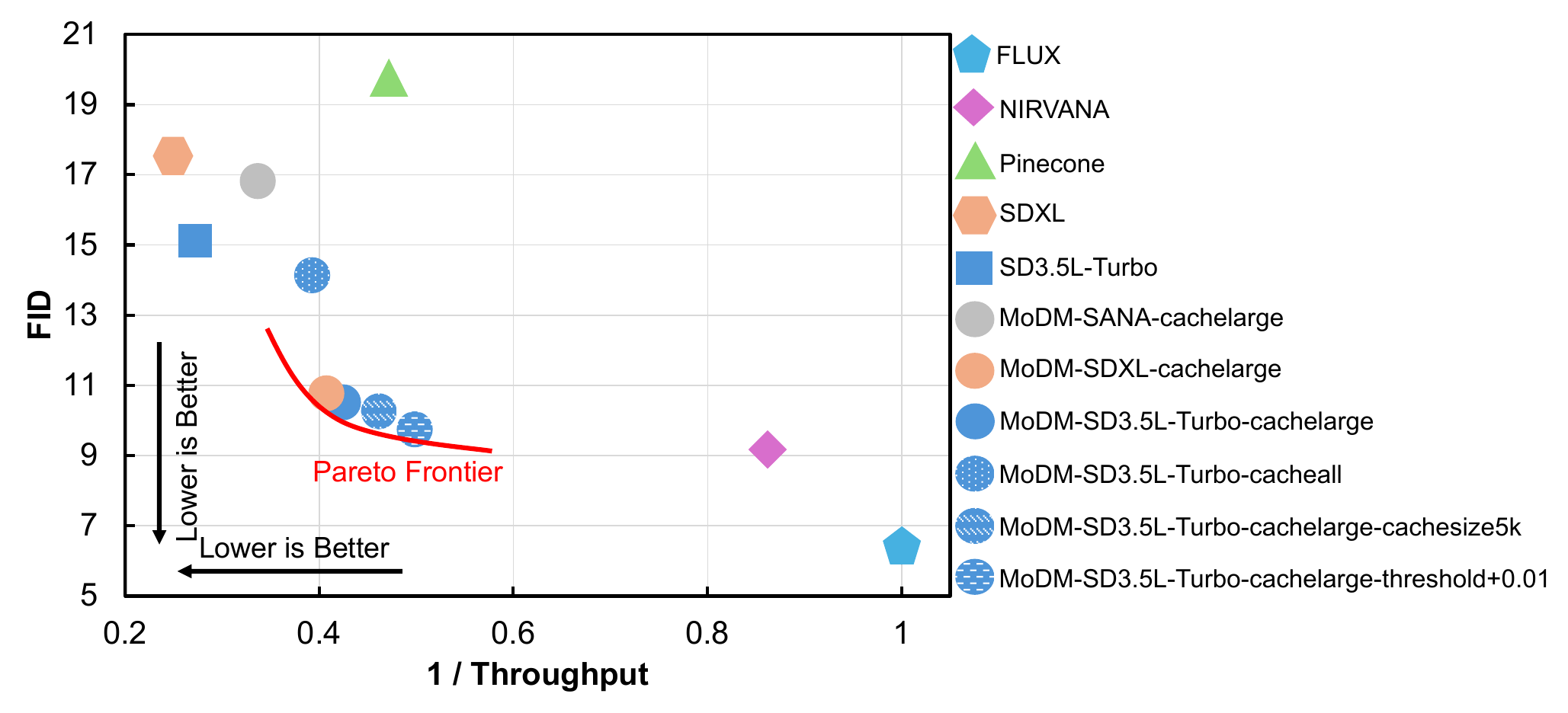}
    \caption{Quality-performance trade-off across serving strategies. The large model used is FLUX, and the evaluation is conducted on the DiffusionDB dataset. The default cache size for all systems is 10k unless otherwise specified.}
    \label{fig:tradeoff}
\end{figure}
\noindent
To complement our main results, Fig.~\ref{fig:tradeoff} visualizes the trade-off space between image generation quality (measured using the FID score) and system throughput for various diffusion-based serving strategies.
We plot the inverse throughput (1/Throughput) on the x-axis and FID score on the y-axis: both axes where lower values are desirable.
This orientation highlights the dual objective in \THISWORK\ of optimizing for faster image generation while maintaining high visual fidelity.

In this figure, we show multiple data points of \THISWORK\ by changing various runtime parameters including (1) small model used: SANA, SDXL, and SD3.5L-Turbo), (2) caching strategies: caching images generated by large model only and caching images generated by all models, (3) cache size: 10k images (default) and 5k images, and (4) varying cache hit threshold: 0.25 (default) and 0.26 (\textit{i.e.,} +0.01).
The figure highlights \textbf{\textit{three key takeaways}}.
First, \THISWORK\ offers a flexible configuration that supports pairing any two diffusion models: not only from the same family but also across different families, including distilled variants, enabling a unique and customizable trade-off between output quality and generation performance.
Second, \THISWORK\ supports runtime adaptivity to system parameters, demonstrating its capability to dynamically adjust and maintain optimal quality-performance balance under varying conditions.
Third, within the landscape of state-of-the-art diffusion-based image generation systems, \THISWORK\ achieves a system design that resides on the \textbf{\textit{Pareto frontier}}, effectively pushing the boundary of the achievable trade-off between quality and performance.
% to use different models and different runtime parameters that leads to offer different design trade-offs.
% Second, 
% Each data point represents a unique model configuration or serving setup. MoDM variants show a range of trade-offs depending on the small model used (e.g., SDXL, SD3.5L-Turbo, SANA) and caching strategies (e.g., full caching, partial caching, threshold tuning). As illustrated, the MoDM configurations lie close to the solid red curve, which represents the Pareto frontier—characterizing the set of optimal trade-offs between FID and throughput achievable under the current configurations.

\subsection{Additional Results}
Due to page limitation, we include additional results in appendix that includes (1) comparison of 99th percentile tail latency of different baselines: \S\ref{sec:tail_latency}, (2) throughput comparison across fluctuating request rates: \S\ref{subsection:throughput_fluctuating}, (3) comparison of energy consumption of different approaches: \S\ref{sec:energy_results}, (4) cache hit rates for the MJHQ dataset: \S\ref{sec:cache_hit_rates_mjhq}, and (5) examples of images generated using different baselines: \S\ref{sec:examples_images}.

\section{Related Works} \label{section:related_work}

% This section situates \THISWORK\ within the established context of existing research.
% This section situates \THISWORK\ within the established context of existing research on caching strategies, adaptive steps and model selection, and resource management techniques tailored for diffusion model serving.

\textbf{Caching Strategies for Diffusion Models.}
Caching strategies accelerate diffusion models by reducing redundant computation.
Works like \textsc{Nirvana}~\cite{nirvama} used latent caching, but it was limited by model-specific constraints and high storage overhead.
Approaches like DeepCache~\cite{deepCache} and PatchedServe~\cite{patchedServe} cached intermediate features and patches, respectively.
\THISWORK\ innovates by caching final generated images along with CLIP embeddings, offering benefits such as cross-model compatibility, reduced storage, and enhanced semantic alignment through text-image similarity.
Unlike SDEdit~\cite{sdedit}, which focuses on guided image regeneration, \THISWORK\ creates new images by leveraging cached content.
EchoLM~\cite{echoLM} maintains a cache of full-model responses and uses cache hits as few-shot prompts to a small LLM for knowledge distillation. \THISWORK\ 
uses an orthogonal technique for diffusion models by using cached final images under a PID-driven scheduler, dynamically partitioning GPU resources and selecting the number of denoising steps.

% Caching strategies have proven effective in accelerating diffusion models by curtailing redundant computation. Various works~\cite{blockCache, patchedServe, nirvama, deepCache, layerCaching, fora, dualCaching, ReCon}, have been proposed in this regard. Initial work by \textsc{Nirvana}~\cite{nirvama} explored latent caching, storing intermediate noise vectors to facilitate partial regeneration. However, latent caching is inherently limited by model-specific constraints and significant storage overhead. The DeepCache~\cite{deepCache}, Layer caching~\cite{layerCaching} and freezing~\cite{freezing} strategies take a different approach by caching intermediate features, layers or outputs~\cite{fora} within the U-Net architecture of the diffusion model while, PatchedServe~\cite{patchedServe} splits images in a predefined patch size and uses cache at patch level. \THISWORK\ diverges from all these approaches by caching the final generated images in conjunction with CLIP~\cite{clipScore} embeddings. Caching final images offers the distinct advantages of cross-model compatibility, substantial storage reduction, and enhanced semantic alignment, achieved through text-image similarity matching. SDEdit~\cite{sdedit}'s image-to-image diffusion techniques for image-based regeneration is an approach which takes inspiration from painting strokes, and is focused on guided image generation from a starting point to an ending point while \THISWORK\ focuses on creating completely new images for incoming requests using cached images.

\textbf{ML optimizations.}
While techniques like progressive distillation~\cite{progressiveDistillation}, structural pruning~\cite{wangPruning}, quantization~\cite{quantDiffusion, lowBitQuantization}, and patching~\cite{patchDiffusion} have been widely adopted in diffusion models, their single-model, training-focused approach lacks the dynamic adaptability needed to handle real-world workload fluctuations effectively. In contrast, \THISWORK's mixture-of-models strategy draws on the strengths of Switch Transformers~\cite{switchTransformers} from language models, adapting them for diffusion processes to better address these challenges.
Flow Matching~\cite{flowMatching} reframes diffusion as a Continuous Normalizing Flow (CNF) trained via vector‐field regression on optimal‐transport–inspired paths, achieving high‐fidelity sampling in under 10 Ordinary Differential Equations (ODE) steps. While these algorithmic advances tackle the modeling side of the latency–quality trade-off, \THISWORK\ addresses the runtime side and can accelerate any diffusion backbone as shown by using SD3.5L, which utilizes the CNF framework internally.

% While optimization techniques like progressive distillation~\cite{progressiveDistillation, distillation}, structural pruning~\cite{wangPruning}, quantization~\cite{quantDiffusion, lowBitQuantization}, patching~\cite{patchDiffusion} have seen widespread adoption in diffusion models, their single-model focus, majorly training oriented lacks the dynamic adaptability required to handle fluctuating real-world workloads effectively. In contrast, \THISWORK's mixture-of-models strategy leverages the strengths of Switch Transformers~\cite{switchTransformers} from the domain of language models, adapting them specifically for diffusion processes.

\textbf{Adaptive steps selection and Query-Aware Model Selection.} 
AdaDiff~\cite{zhang2023adadiff} proposes selecting adaptive denoising steps based on prompt richness, while other works~\cite{earlyExit,fastDPM,sceEdit-skipping} explore early exit, fast sampling, or step-skipping techniques. A key innovation of \THISWORK\ is using a cached image as a starting point, combined with a dynamic $k$-selection mechanism that adjusts the number of denoising steps based on CLIP~\cite{clipScore} scores and predefined quality constraints.
Diff Serve~\cite{diffServe} is a query-aware approach for selecting diffusion models, employing a cascading strategy of light and heavy models with a discriminator-based method. This makes it an orthogonal approach to \THISWORK.

% AdaDiff\cite{zhang2023adadiff} proposes selecting adaptive denoising steps depending on prompt richness, while other works ~\cite{earlyExit,fastDPM, sceEdit-skipping} recommend early exit~\cite{earlyExit},fast sampling~\cite{fastDPM} or skipping~\cite{sceEdit-skipping} steps techniques. A key innovation of our work is using a cached image as a starting point with dynamic k-selection mechanism, which modulates the number of denoising steps based on CLIP~\cite{clipScore} scores and predefined quality constraints.

% \textbf{Query Aware Model Selection:}
% Diff Serve~\cite{diffServe} is one approach we discovered for query aware selection of Diffusion Models which employs cascading of light-heavy models using a discriminator-based approach making it an orthogonal approach to \THISWORK.

\textbf{Cloud Serving Infrastructure.}
\THISWORK\ builds on resource management from distributed systems and AI scheduling, incorporating a PID-controlled resource allocator with diffusion-specific metrics. While Kubernetes-based autoscalers~\cite{clipper, cocktail, pretzel, yadwadkar2019case, MARKforSLO} focus on latency reduction, they lack diffusion model awareness, making them complementary to \THISWORK\ and easy to integrate. 
Inspired by DNN serving systems~\cite{servingClockwork, servingFNF, servingSmartLite} and cache-aware LLMs~\cite{cacheAwareQA}, \THISWORK\ introduces cache-aware queue resource distribution.

% Building upon established resource management principles from distributed systems and AI workload scheduling, \THISWORK's system incorporates a PID-controlled resource allocator with diffusion-specific metrics. Past research~\cite{clipper, cocktail, pretzel, yadwadkar2019case, MARKforSLO} with Kubernetes~\cite{kubernetes}-based autoscalers focus on reducing inference latency but lack awareness of diffusion model dynamics, making them complementary to \THISWORK\ and easily integrable. Inspired by DNN serving systems~\cite{servingClockwork, servingFNF, servingSmartLite}, and cache awareness in LLMs~\cite{cacheAwareQA}, \THISWORK\ introduces a cache-aware queue management for Diffusion Models serving, that significantly reduces latency. 

% \THISWORK's cache maintenance protocol combines LRU eviction~\cite{lru} with image usefulness awareness, addressing the limitations of traditional methods. This end-to-end architecture optimizes GPU utilization while maintaining service level objectives, demonstrating the effectiveness of coordinated resource management in diffusion model serving.

\vspace{-0.1cm}

\section{Conclusions} \label{section:conclusion}
This paper introduced \THISWORK, an efficient serving system for diffusion-based text-to-image generation.
\THISWORK\ \textit{dynamically balances between inference latency and image quality} by leveraging a \textit{mixture of diffusion models}.
The key innovation is its cache-based approach: when a new request closely matches a previously generated image, the system retrieves the cached image, reintroduces noise, and refines/de-noises it using a small model, while cache-miss requests are handled by a large model.
By strategically combining small and large models, \THISWORK\ achieves over 2$\times$ higher throughput, ideal scalability with GPU resources, and superior load handling without violating SLO, all while maintaining image quality comparable to large-model inference at a fraction of the cost.

\section{Acknowledgments}
This work is supported in part by Semiconductor Research Corporation (SRC) as part of the GRC AIHW grant, and AMD AI \& HPC Cluster Program.

% \clearpage
\balance
%%%%%%%%% -- BIB STYLE AND FILE -- %%%%%%%%
% \bibliographystyle{ACM-Reference-Format}
% \input{main.bbl}
\bibliography{99_ref.bib}
\bibliographystyle{unsrt} %To see references in the order they are in the paper

\clearpage
\appendix

\section{Appendix}

\subsection{Timing Analysis Between a New Prompt and its Retrieved Cache Image} \label{sec:timing_analysis}
\begin{figure}[H]
    \centering

    \includegraphics[width=0.9\linewidth]{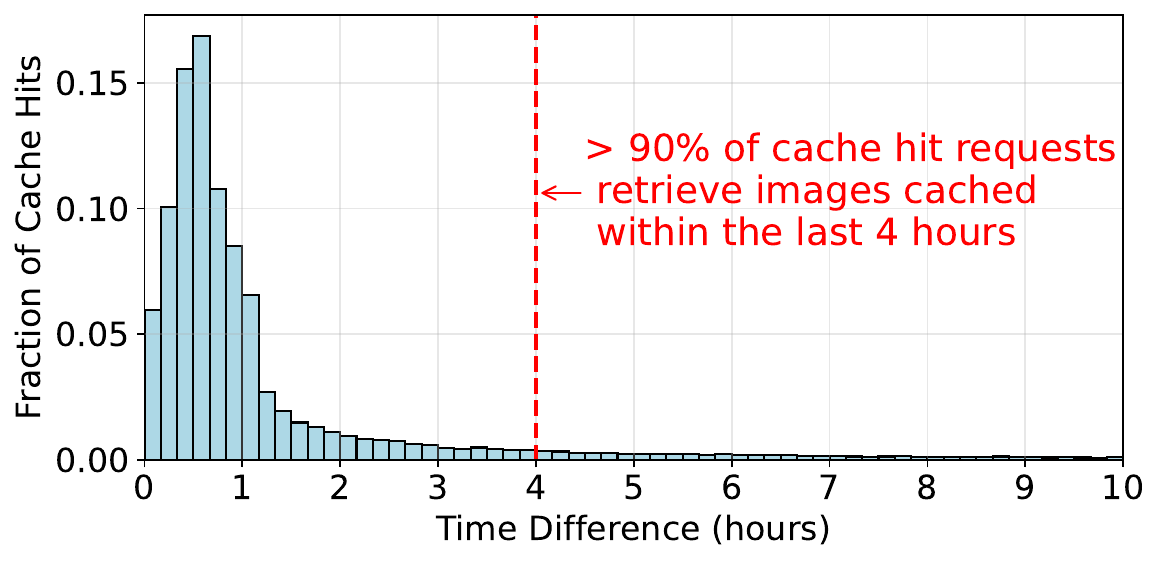}
    \caption{Distriution of time elapsed between new requests and the generation of their retrieved images from the cache.}
    \label{fig:time_difference}
\end{figure}
\noindent
To evaluate the effectiveness of a simple FIFO-based cache management strategy, we conduct an experiment using the production dataset DiffusionDB~\cite{wang2022diffusiondb}, which accurately captures the temporal correlation between different prompts.  
Specifically, we measure the time elapsed between a new prompt that results in a cache hit and the original image generation of its retrieved cached item.  
Fig.~\ref{fig:time_difference} presents the distribution of these time intervals, showing that over 90\% of new prompts retrieve images generated within the past four hours.  
In other words, caching all requests from the last four hours can achieve a high cache hit rate of over 90\%, making it feasible to ignore images generated much earlier.  
This behavior is intuitive, as users often iteratively refine their prompts to better align the generated visual content with their expectations.  
Based on this quantitative analysis, \textit{\THISWORK\ adopts a simple yet effective FIFO-based cache maintenance strategy.}
% For evaluating the effectiveness of a simple FIFO-based cache management strategy, we conduct an experiment using a production dataset DiffusionDB~\cite{wang2022diffusiondb} that correctly captures the temporal correlation between different prompts.
% Specifically, we measure the time elapsed between a new prompt that hits in the cache and the image generation of its cached item that it retrieves.
% Fig.~\ref{fig:time_difference} shows the distribution of this time indicating that more than 90\% of new prompts retrieve images from the cache that were generated four hours before.
% In other words, by caching all requests within the last four hours can result in a high cache hit rate of more than 90\%, and it is possible to ignore the images generated too far back in time.
% This is intuitive because many users prompt diffusion models with similar prompts multiple times to best align the visual content with their expectations.
% Based on this quantitative analysis, \THISWORK\ makes a design decision to employ a simple, yet effective FIFO-based cache maintenance strategy.

\subsection{Tail Latency Evaluation} \label{sec:tail_latency}
\begin{figure}[h]
    \centering
    \includegraphics[width=\linewidth]{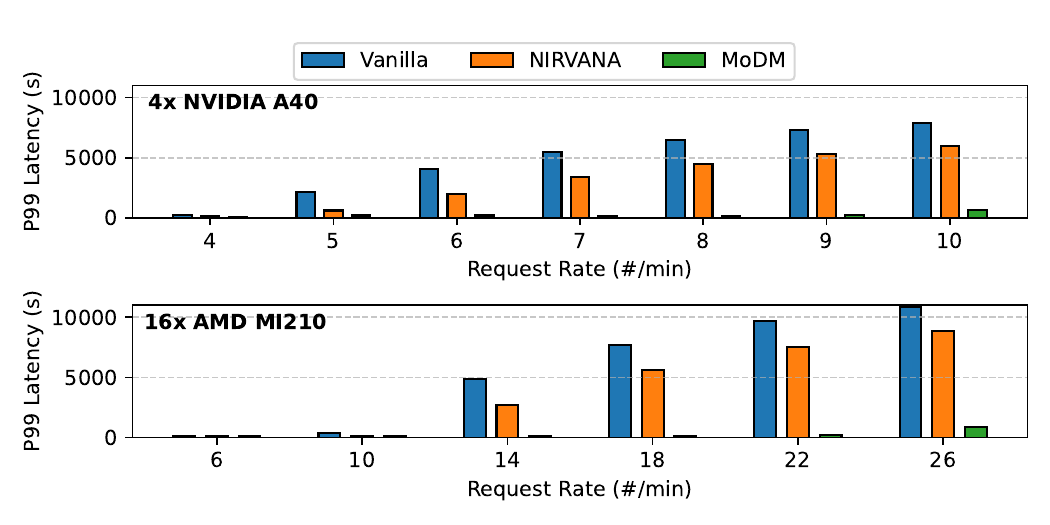}
    % \vspace{-1cm}
    \caption{P99 tail latency for varying request rates.}
    \label{fig:latency}
\end{figure}
\noindent
Fig.~\ref{fig:latency} demonstrates that \THISWORK\ significantly reduces the 99th percentile tail latency compared to the vanilla system and \textsc{Nirvana}.
The upper subfigure compares tail latency using 4 A40s.
At 4 requests per minute, all three systems maintain a low tail latency of under 200 seconds.
However, as the request rate increases from 4 to 10 requests per minute, the tail latency of the vanilla system and \textsc{Nirvana} surges past 1000 seconds, making them impractical for real-time serving.
In contrast, \THISWORK\ can handle significantly higher system loads, supporting up to 10 requests per minute with the given GPU resources. Due to the compute-heavy nature of diffusion model serving, reducing latency beyond this rate requires a substantial increase in GPU resources.

% In contrast, \THISWORK\ is able to tolerate a much higher system load, \textit{i.e.,} up to 10 requests per minute given these GPU resources.
% Due to the compute-heavy nature of diffusion model serving, a large number of GPU resources are necessary to reduce latency beyond this request rate.
% This indicates that both vanilla and \textsc{Nirvana} fail to sustain request rates beyond 5 requests per minute, whereas MoDM maintains a consistently low tail latency until the request rate reaches 10 requests per minute, marking the threshold beyond which it starts to degrade.

The lower subfigure in Fig.~\ref{fig:latency} presents tail latency with 16 MI210s.
Similarly, both the vanilla system and \textsc{Nirvana} can only sustain a low tail latency when the request rate does not exceed 10 requests per minute.
In contrast, \THISWORK\ remains stable at much higher request rates, successfully handling over 20 requests per minute, further demonstrating its robustness and scalability under increasing load.

\subsection{Throughput Under Fluctuating Request Rates} \label{subsection:throughput_fluctuating}
\begin{figure}[h]
    \centering
    \includegraphics[width=\linewidth]{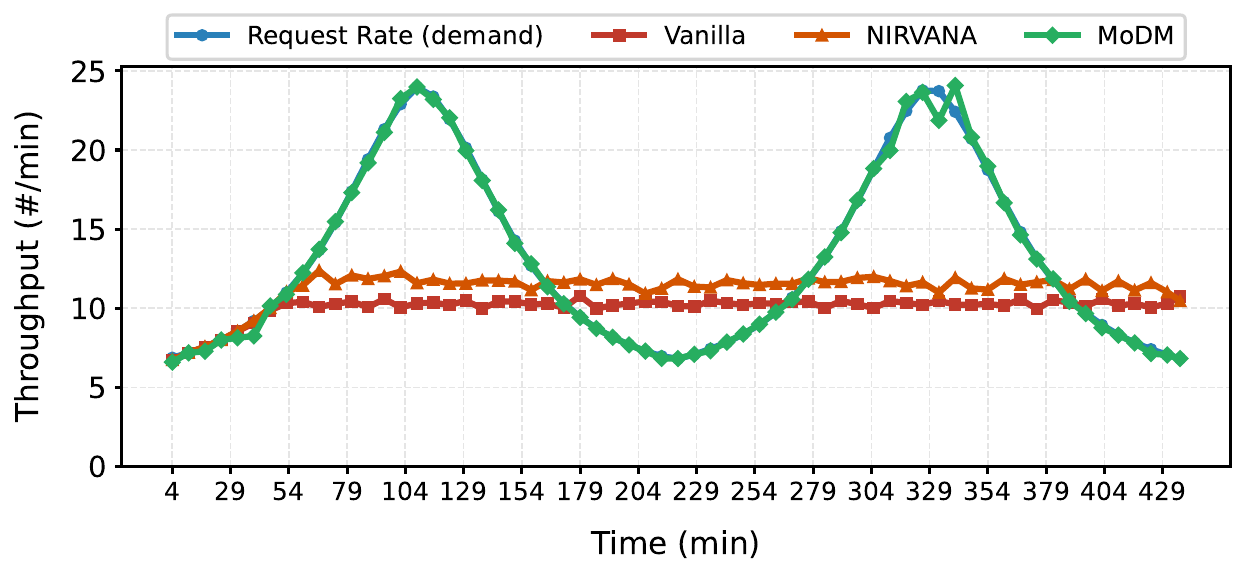}
    \vspace{-0.8cm}
    \caption{Throughput over time under fluctuating request rates.}
    \label{fig:throughput_climbing_rate_f}
\end{figure}
\noindent
Fig.~\ref{fig:throughput_climbing_rate_f} illustrates how different systems respond to varying load conditions over time. As the request rate increases and decreases, \THISWORK\ consistently adapts to match the demand, achieving higher throughput across both low and high load periods. In contrast, baseline systems such as \textsc{Vanilla} and \textsc{Nirvana} show noticeable lag during peak request intervals, indicating limited scalability. Notably, their throughput remains high during low request rates because they are still draining queued requests from earlier peak periods.
These results highlight the effectiveness of our design in maintaining high throughput even under rapidly changing workload patterns.

\subsection{Energy Savings} \label{sec:energy_results}
\begin{figure}[h]
    \centering
    \includegraphics[width=0.9\linewidth]{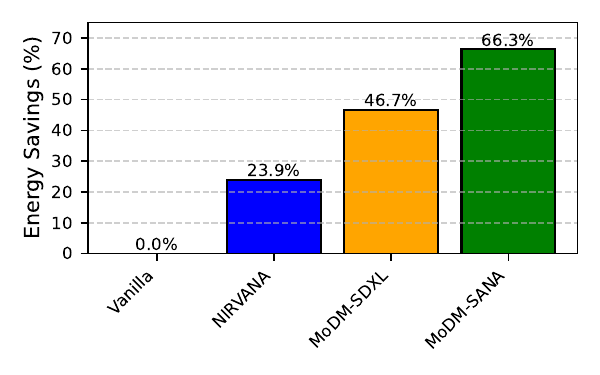}
    \caption{Energy Savings of different baselines normalized to Vanilla (Stable Diffusion-3.5-Large) on DiffusionDB.}
    \label{fig:energySavings}
\end{figure}

We measure the energy consumption of various baselines using Zeus~\cite{zeus}, a Python-based energy measurement toolkit.  
The proposed methods, \THISWORK-SDXL and \THISWORK-SANA, are compared against two references: a standard SD3.5-Large (Vanilla) model and \textsc{Nirvana}.  
Fig.~\ref{fig:energySavings} presents the energy savings of different systems relative to the vanilla baseline.  
\textsc{Nirvana} achieves a modest 23.9\% energy improvement, primarily due to skipping de-noising steps. However, these benefits are limited as inference still relies on a single, large model.
In contrast, \THISWORK\ significantly enhances energy efficiency: (1) \THISWORK-SDXL, which utilizes the SDXL model for cache-hit requests, achieves a 46.7\% energy savings, and (2) \THISWORK-SANA, which leverages the smaller SANA-1.6B model for cache-hit requests while maintaining comparable image quality, achieves even greater efficiency, reaching 66.3\% energy savings.  
These results underscore two key insights.  
First, caching image generations from a large base model effectively reduces redundant computational overhead.  
Second, using lighter, more efficient downstream diffusion models (\textit{e.g.,} SANA-1.6B) further amplifies energy savings.

% We measure the energy consumption of various baselines using Zeus~\cite{zeus}, a Python-based energy measurement toolkit.
% We compare the proposed methods \THISWORK-SDXL and \THISWORK-SANA against two references: a standard Stable Diffusion-3.5 Large (Vanilla) model serving, and \textsc{Nirvana}.
% Fig.~\ref{fig:energySavings} compares the energy saving of different systems compared to vanilla.
% \THISWORK\ achieves a modest energy improvement of 23.9\%. 
% This is due to skipping of de-noising steps; their benefits are limited due to the use of a single, large model for inference.
% \THISWORK, on the other hand, significantly improves energy savings: \THISWORK-SDXL, which uses the SDXL model for cache hit requests delivers a 46.7\% energy savings.
% \THISWORK-SANA, leveraging the SANA-1.6B model for cache hit rates, which is smaller than the SDXL model with similarly cached images, shows even greater effectiveness, achieving 66.3\% energy saving.
% These results highlight two key insights.
% First, that caching image generations from a large base model substantially reduces redundant computational overheads.
% Second, selecting lighter, more efficient downstream diffusion models (\textit{e.g.,} SANA-1.6B) leads to even greater energy savings.

\subsection{Cache Hit Rate Comparison on MJHQ Dataset} \label{sec:cache_hit_rates_mjhq}
\begin{figure}[h]
    \centering
    \includegraphics[width=\linewidth]{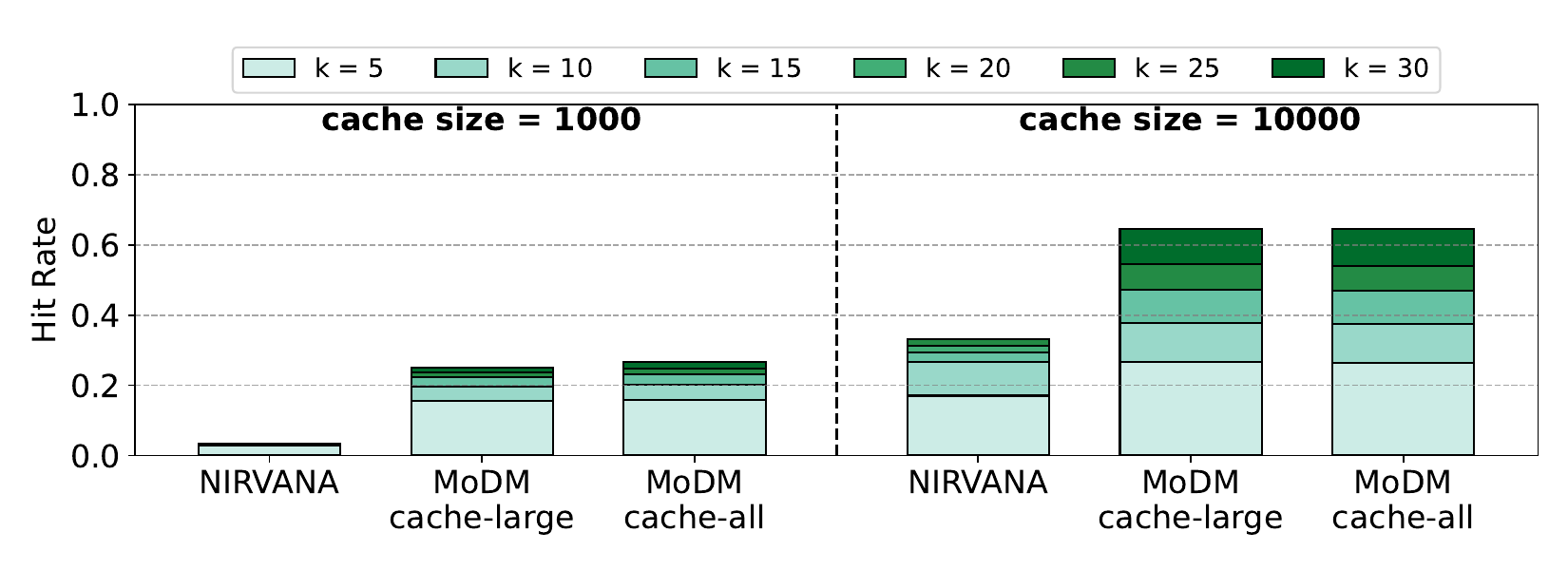}
    \vspace{-0.8cm}
    \caption{Comparison of cache hit rates for \textsc{Nirvana} and \THISWORK\ for the MJHQ dataset~\cite{mjhq_dataset}. A cache size of 100k is not shown as the dataset only has 30k prompts.}
    \label{fig:cache_hitrate_comparison_MJHQ}
\end{figure}
\noindent
Fig.~\ref{fig:cache_hitrate_comparison_MJHQ} compares the cache hit rates of \textsc{Nirvana} and \THISWORK\ on the MJHQ dataset~\cite{mjhq_dataset}.  
The figure evaluates hit rates for two cache sizes: 1000 and 10000 images. A cache size of 100000 is not feasible for this dataset, as it contains only 30000 prompts.  
Unlike DiffusionDB, MJHQ is not a production dataset, meaning there is no strong temporal correlation between requests.  
Despite this, Fig.~\ref{fig:cache_hitrate_comparison_MJHQ} demonstrates that \THISWORK\ consistently improves cache hit rates compared to \textsc{Nirvana}.

Interestingly, the results reveal minimal difference between caching only images from the large model and caching all images.
This is because the dataset lacks temporal proximity of similar prompts, making the inclusion of all images in the cache less beneficial.
Without temporal correlation, caching images generated by a small or a large model provides similar benefits.
\textit{This suggests that when prompts submitted in close temporal proximity lack significant similarity, caching only images generated by the large model (cache miss requests in \THISWORK) is sufficient.}  
Nonetheless, \THISWORK’s retrieval and management policies effectively reduce computational overhead, leading to substantial improvements in overall throughput (\S\ref{sec:throughput_result}).
% Fig.~\ref{fig:cache_hitrate_comparison_MJHQ} compares the cache hit rates of \textsc{Nirvana} and \THISWORK\ for the MJHQ dataset~\cite{mjhq_dataset}.
% The figure compares hit rates for two cache sizes: 1000 images and 10000 images.
% A cache size of 100000 is not possible for this dataset as it only has 30000 prompts.
% Notably, unlike the DiffusionDB dataset, this MJHQ dataset is not a production dataset, implying that there is no tight temporal correlation between requests.
% Despite this, Fig.~\ref{fig:cache_hitrate_comparison_MJHQ} shows that \THISWORK\ consistently improves the cache hit rate compared to \textsc{Nirvana}.
% Interestingly, this comparison reveals that there is little difference between caching images from large model only or caching all images.
% This is because the requests in this dataset do not have temporal proximity benefit of adding all images to the cache.
% With no temporal correlation, caching an image generated by small and large models have the same effect.
% \textit{This indicates that if the prompts submitted in temporal proximity do not have a high degree of similarity, only adding images generated by a large model to the cache is sufficient.
% }Despite this, \THISWORK\ retrieval and management policies can effectively reduce a large portion of computational overhead, leading to substantial improvement in overall throughput (\S\ref{sec:throughput_result}).

\subsection{Effects of Caching Images Generated by the Small Model} \label{sec:cache_small}
In this section, we evaluate the impact of caching images generated by the small model based on retrieved previously generated images.  
Specifically, we aim to answer the following question: \textbf{Does adding images generated by the small model to the cache preserve the quality of future image generations?}
To evaluate the impact of adding cache-hit images to the cache for subsequent retrieval and reuse, we conduct the following experiment.
First, we generate 10000 images using Stable Diffusion-3.5 Large (SD3.5L) to warm up the cache.
Then, we process another 10000 new requests, attempting to retrieve and reuse images from the cache.
If a request results in a cache hit, we proceed with three different generation strategies.
The first strategy is to perform full generation, where the cache-hit request is processed by SD3.5L to generate an entirely new image.
The second strategy is to refine the retrieved image using SD3.5L.
The third strategy is to refine the retrieved image using SDXL, a smaller and more efficient model.

This process results in three sets of newly generated cache-hit images: one generated directly by SD3.5L without any caching (the golden standard), one refined using retrieved cache-hit images by SD3.5L, and one refined using retrieved cache-hit images by SDXL.
We then use these three cache sets separately to serve another 10000 new requests, generating all images using SDXL.
Finally, we evaluate the quality of the final generated cache hit images using CLIP score.
Our evaluation shows that using directly generated SD3.5L images as cache achieves a CLIP score of 29.63.
Using cache-hit images refined by SD3.5L achieves a CLIP score of 28.58, while using cache-hit images refined by SDXL results in a CLIP score of 28.32.
The drop in CLIP score is minimal, demonstrating that the images generated by reusing cache-hit images refined by both models maintain high text-image alignment.
Considering the boost in cache hit rate by adding these images (more details in \S\ref{sec:throughput_result}, Fig.~\ref{fig:cache_policy_comparison}), this approach proves to be a valuable strategy for improving system efficiency while maintaining image quality.
\textit{Based on this quantitative analysis, we make a design decision to cache all images (generated by both small and large models) without degrading the quality of future image generation.}

\subsection{Examples of Image Generation} \label{sec:examples_images}

\begin{figure*}[h]
    \centering
    \renewcommand{\arraystretch}{1.2}
    \setlength{\tabcolsep}{2pt} % Reduce column spacing
    \small
    \begin{tabular}{c c c c c c}
        \hline
        \textbf{Request Prompt} & \textbf{SD3.5L} & \textbf{SDXL} & \textbf{SANA} & \textbf{\THISWORK-SDXL} & \textbf{\THISWORK-SANA} \\
        \hline
        \raisebox{7ex}{\parbox{4cm}{\centering Carlos Sainz Jr.}} &
        \includegraphics[width=0.13\textwidth]{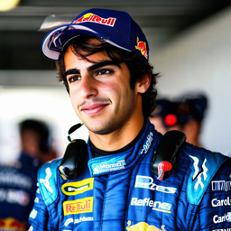} & 
        \includegraphics[width=0.13\textwidth]{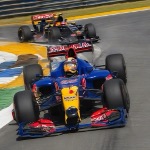} & 
        \includegraphics[width=0.13\textwidth]{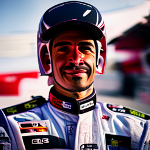} &
        \includegraphics[width=0.13\textwidth]{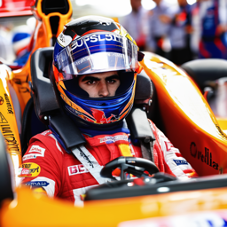} & 
        \includegraphics[width=0.13\textwidth]{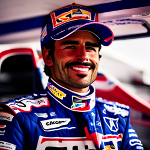} \\

        \raisebox{7ex}{\parbox{4cm}{\centering Horseshoe Bend in Utah Cinematic National Geographic Photoshoot }} &
        \includegraphics[width=0.13\textwidth]{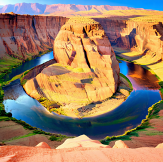} & 
        \includegraphics[width=0.13\textwidth]{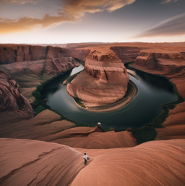} & 
        \includegraphics[width=0.13\textwidth]{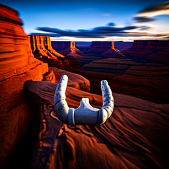} &
        \includegraphics[width=0.13\textwidth]{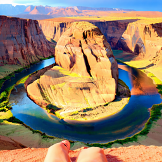} & 
        \includegraphics[width=0.13\textwidth]{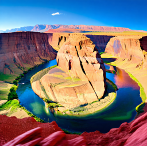} \\

        \raisebox{8ex}{\parbox{4cm}{\centering Batman holding up a bitcoin}} &
        \includegraphics[width=0.13\textwidth]{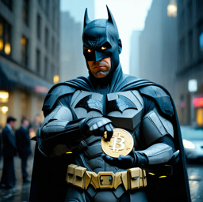} & 
        \includegraphics[width=0.13\textwidth]{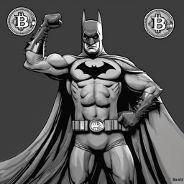} & 
        \includegraphics[width=0.13\textwidth]{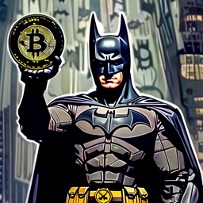} &
        \includegraphics[width=0.13\textwidth]{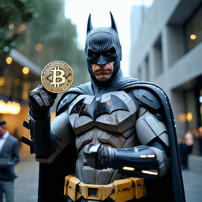} & 
        \includegraphics[width=0.13\textwidth]{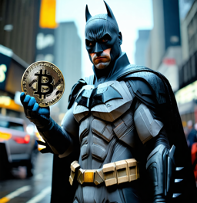} \\

        \raisebox{7ex}{\parbox{4cm}{\centering Joy of human and dog interation}} &
        \includegraphics[width=0.13\textwidth]{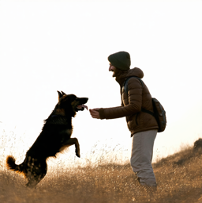} & 
        \includegraphics[width=0.13\textwidth]{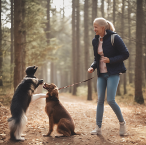} & 
        \includegraphics[width=0.13\textwidth]{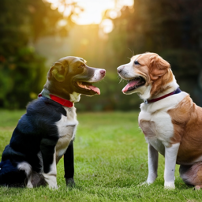} &
        \includegraphics[width=0.13\textwidth]{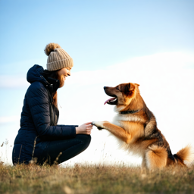} & 
        \includegraphics[width=0.13\textwidth]{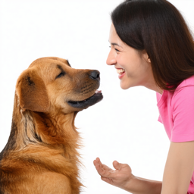} \\

       \raisebox{8ex}{\parbox{4cm}{\centering A sneaker event with a DJ performing in the middle of the crowd within foot locker shop new balance sneakers on display photorealistic}} &
        \includegraphics[width=0.13\textwidth]{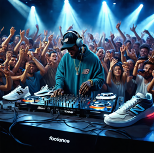} & 
        \includegraphics[width=0.13\textwidth]{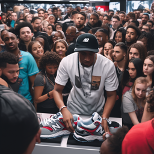} & 
        \includegraphics[width=0.13\textwidth]{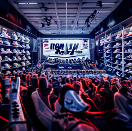} &
        \includegraphics[width=0.13\textwidth]{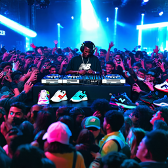} & 
        \includegraphics[width=0.13\textwidth]{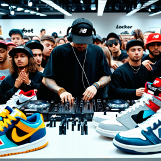} \\

        \raisebox{7ex}{\parbox{4cm}{\centering A village over a tree in atamas dessert with a lot of vegetation in the night blue lighting with HDR cinematic look Hyperrealistic}} &
        \includegraphics[width=0.13\textwidth]{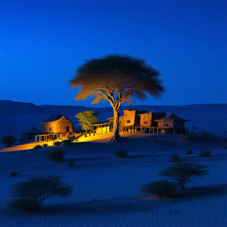} & 
        \includegraphics[width=0.13\textwidth]{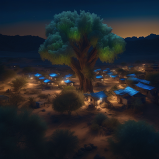} & 
        \includegraphics[width=0.13\textwidth]{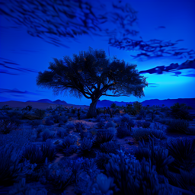} &
        \includegraphics[width=0.13\textwidth]{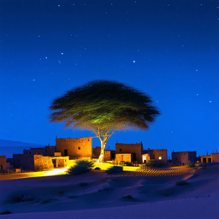} & 
        \includegraphics[width=0.13\textwidth]{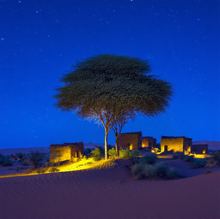} \\

        % \raisebox{7ex}{\parbox{4cm}{\centering A wine glass full of whipped cream and frogs}} &
        % \includegraphics[width=0.13\textwidth]{Images_Vanilla/6.png} & 
        % \includegraphics[width=0.13\textwidth]{Images_SDXL/6.png} & 
        % \includegraphics[width=0.13\textwidth]{Images_SANA/6.png} &
        % \includegraphics[width=0.13\textwidth]{Images_MoDM_SDXL/6.png} & 
        % \includegraphics[width=0.13\textwidth]{Images_MoDM_SANA/6.png} \\ 

        \raisebox{8ex}{\parbox{4cm}{\centering A woman sitting on a black sofa next to her French Bulldog. The French Bulldog is white. The walls of the room are glazed. the room has a view to the garden}} &
        \includegraphics[width=0.13\textwidth]{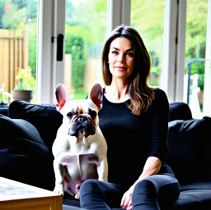} & 
        \includegraphics[width=0.13\textwidth]{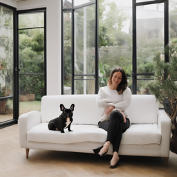} & 
        \includegraphics[width=0.13\textwidth]{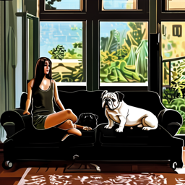} &
        \includegraphics[width=0.13\textwidth]{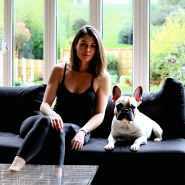} & 
        \includegraphics[width=0.13\textwidth]{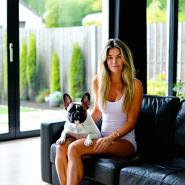} \\

        \raisebox{7ex}{\parbox{4cm}{\centering Airport cartoon style pastel}} &
        \includegraphics[width=0.13\textwidth]{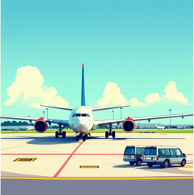} & 
        \includegraphics[width=0.13\textwidth]{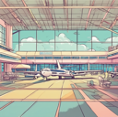} & 
        \includegraphics[width=0.13\textwidth]{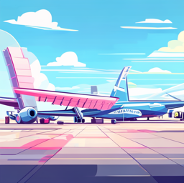} &
        \includegraphics[width=0.13\textwidth]{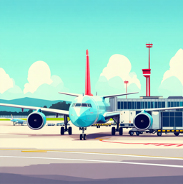} & 
        \includegraphics[width=0.13\textwidth]{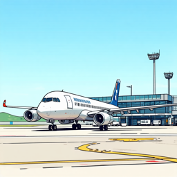} \\ 

        \hline
    \end{tabular}
    \caption{Generated images for different methods on 8 sample requests. \THISWORK\ uses SD3.5L as a large model.}
    \label{fig:appendix_images}
\end{figure*}

Fig.~\ref{fig:appendix_images} presents examples of images generated using different serving systems: (1) \textit{SD3.5L}: a large model, Stable Diffusion-3.5-Large (referred to as vanilla in the main text), (2) \textit{SDXL}: a relatively smaller model, Stable Diffusion-XL, (3) \textit{SANA}: a small model, SANA-1.6B~\cite{xie2024sana}, (4) \textit{\THISWORK-SDXL}: the proposed system using SD3.5L as the large model and SDXL as the small model, and (5) \textit{\THISWORK-SANA}: the proposed system with SD3.5L as the large model and SANA as the small model.  
The prompts are selected from both the DiffusionDB~\cite{wang2022diffusiondb} and MJHQ~\cite{mjhq_dataset} datasets, and \THISWORK\ utilizes cached images from previous generations to produce new outputs using a small model.
The figure illustrates that the image quality of smaller models, SDXL and SANA, is sometimes inferior to that of the larger SD3.5L model.  
For instance, for the prompt \textit{"A woman sitting..."}, SDXL generates an image with incorrect colors of the sofa and French Bulldog.
Similarly, for the prompt \textit{"Joy of human..."}, SANA fails to include a human in the image. 
In contrast, \THISWORK\ maintains image quality by accurately preserving content while benefiting from the reduced latency of small model inference.

\subsection{Frequently Asked Questions (FAQs)}
In this section, we clarify a few frequently asked questions about our design choices.

\noindent
\textbf{Q.1. Why does \THISWORK\ use small and large models?}\\
\textbf{Response:} 
This approach \textit{balances inference latency and image generation quality}.
Specifically, we observe that caching images generated by large models and using them as a starting point for small models to serve prompts with similar visual intentions leads to excellent image quality.
\S\ref{sec:generation_using_cache} explains how to generate an image at a reduced cost by utilizing a cached image.
The quality degradation is minimal, and in some cases, it even outperforms a standalone model without caching, all while significantly reducing compute and latency.
Therefore, the proposed mixture of models approach combines the image quality of large models with the lower latency of small models.

% This is to balance the inference latency and image generation quality.
% Specifically, we observe that caching and using images generated from large models as a starting point for small models to serve prompts with similarly intended visual output (as cached images) performs quite well on image quality (very little drop or sometimes even better than a standalone model with no caching) while giving significant compute and latency reduction.

\noindent
\textbf{Q.2. How does \THISWORK\ maintain compatibility with different model families?}\\
\textbf{Response:} 
This is made possible by our caching strategy, which caches \textit{final generated images} rather than intermediate features. 
Final images are more versatile and recognizable across different models and model families. 
\S\ref{sec:model_interoperability} explores model-agnostic caching for serving across multiple model families.
% Our caching relies on text-to-image similarity (CLIP scores), which measures semantic similarity between text and final image independently from model architecture, enabling cross-family applicability.

\noindent
\textbf{Q.3. What does \THISWORK\ cache?}\\
\textbf{Response:} 
We cache \textit{final generated images}. 

\noindent
\textbf{Q.4. Why do we cache final images and not latent intermediate?}\\
\textbf{Response:}
Final images are directly usable and \textit{model independent}, making them universally compatible across all model families.
In contrast, intermediate latents vary between models, limiting serving to a single model.

\noindent
\textbf{Q.5. How is our approach similar or different from speculative decoding in LLMs?}\\
\textbf{Response:} 
Speculative decoding differs from caching-based image generation in \THISWORK.
In speculative decoding, a small draft model predicts tokens for text generation, while a large verification model checks and refines them.
In contrast, \THISWORK\ does not involve verification by a large model.
Instead, each prompt is processed by either the small or large model, depending on cache availability: ensuring efficiency without additional verification overhead.

\noindent
\textbf{Q.6. How well do cross-model queries work in \THISWORK?}\\
\textbf{Response:}
We demonstrate cross-model compatibility using two model families: Stable Diffusion and SANA.  
\S\ref{fig:throughput} presents overall throughput and how \THISWORK\ utilizes different models to handle high request loads.  
\S\ref{sec:image_quality} evaluates the image quality produced by different model families.  
\S\ref{sec:examples_images} provides visual examples of images generated using a mix of models from different families.
% Cross-model queries perform effectively well due to the semantic nature of CLIP similarity scores, and using final images as cache which transcend specific model architectures.

\noindent
\textbf{Q.7. How are the thresholds on $k$ decided?}\\
\textbf{Response:} 
\S\ref{sec:cache_retrieval} explains how thresholds on \( k \) are determined using text-to-image similarity scores, ensuring a high image quality factor of \( \geq 0.95 \).

\noindent
\textbf{Q.8. Where is the main performance benefit of \THISWORK\ coming from?}\\
\textbf{Response:}
The significant performance uplift of \THISWORK\ is driven by two key factors: (1) skipping a subset of de-noising/refinement steps by leveraging cached items, and (2) utilizing a small model for inference when a cache hit occurs.

\noindent
\textbf{Q.9. Does \THISWORK\ always use a small model to serve a requests that hit in the cache?}\\
\textbf{Response:}
To maximize system throughput, \THISWORK\ defaults to using a small diffusion model for all requests hitting in the cache (throughput-optimized mode, \S\ref{sec:global_monitor}).
Additionally, \THISWORK\ offers flexibility for service providers to prioritize image quality by serving cache hits with a large model \textbf{when request rates are low and SLO requirements allow} (quality-optimized mode, \S\ref{sec:global_monitor}).
Fig.~\ref{fig:throughput_climbing_rate} illustrates this use-case, showing that when request rates drop below 10 per minute, cache hits can be served by the large model to maximize image quality without violating SLO.
However, \THISWORK\ can also run in throughput-optimized mode at low request rates if preferred.
% To maximize the throughput of the overall system, \THISWORK\ employs a small model for serving all requests using a small diffusion model (throughput-optimized mode in \S\ref{sec:global_monitor}).
% In addition, we offer a flexibility to the service provider that also allows maximizing image generation quality by serving cache hit requests through a large model whenever the request rate is low and SLO requirement can be met using a large model (quality-optimized model in \S\ref{sec:global_monitor}).
% Fig.~\ref{fig:throughput_climbing_rate} showcases one such use-case, where cache hit requests can also be served using a large model when the request rate is less than 10 requests per minute, which maximizes the quality of generated image.
% However, if the service provider desires, \THISWORK\ can be set to run in throughput-optimized mode as well with low request rates.

\noindent
\textbf{Q.10. How does \THISWORK\ maintain image generation diversity?}\\
\textbf{Response:}
While \THISWORK\ generates images that hit in the cache by reusing previously generated outputs—through controlled noise injection and partial denoising—a key design choice is the adoption of a FIFO-based caching strategy.
Unlike utility-based policies (\textit{e.g.,} those inspired by CPU hardware caches), the FIFO-based approach ensures automatic eviction of cached images after a fixed time window.
This prevents a small set of highly popular cached images from dominating reuse, thereby encouraging diversity in the cache and maintaining adaptability to evolving input distributions.
A quantitative evaluation of generation diversity is a compelling direction for future work.

\section{Artifact Appendix}

%%%%%%%%%%%%%%%%%%%%%%%%%%%%%%%%%%%%%%%%%%%%%%%%%%%%%%%%%%%%%%%%%%%%%
\subsection{Abstract}

This artifact accompanies our paper \textit{MoDM: Efficient Serving for Image Generation via Mixture-of-Diffusion Models}, which introduces MoDM, a novel caching-based serving system for text-to-image diffusion models that balances latency and image quality by dynamically leveraging a mixture of large and small diffusion models.

The artifact provides detailed instructions for environment setup, dataset preparation, and running experiments to reproduce our main results. In particular, we focus on reproducing:
\begin{itemize}
    \item The \textbf{throughput comparisons} shown in Figure~7, demonstrating MoDM’s significant improvements over baseline systems under different workloads.
    \item The \textbf{image quality comparisons} summarized in Table~2.
\end{itemize}

By following our artifact, reviewers will be able to replicate our experimental pipeline, observe MoDM’s dynamic scheduling of diffusion models, and confirm the throughput-quality trade-offs across different datasets.

\subsection{Artifact check-list (meta-information)}

% {\em Obligatory. Use just a few informal keywords in all fields applicable to your artifacts
% and remove the rest. This information is needed to find appropriate reviewers and gradually 
% unify artifact meta information in Digital Libraries.}

{\small
\begin{itemize}
  \item {\bf Algorithm:} Diffusion-based text-to-image generation
  \item {\bf Program:} Python 3
  \item {\bf Compilation:} Not required; all components are interpreted
  \item {\bf Models:} Stable Diffusion 3.5-Large, Stable Diffusion XL and SANA-1.6B
  \item {\bf Data sets:} DiffusionDB and MJHQ-30k
  \item {\bf Run-time environment:} Python environment with required libraries (provided via \texttt{requirements.txt})
  \item {\bf Hardware:} 4 or more GPUs with at least 40GB VRAM each
  \item {\bf Run-time state:} Initializes with a cache of 10,000 images
  \item {\bf Execution:} Processes 1K requests from each dataset to evaluate throughput and image quality
  \item {\bf Metrics:} Throughput (reqs/min), CLIPScore, PickScore, Inception Score (IS)
  \item {\bf Output:} Generated images under the \texttt{images} directory and throughput reports in text files for each baseline on both datasets
  \item {\bf Experiments:} Compare MoDM against baselines on throughput and image quality using the same workload and cached dataset
  \item {\bf Disk space required:} 100GB
  \item {\bf Time to prepare workflow:} 1 hour
  \item {\bf Time to complete experiments:} 16 hours
  \item {\bf Publicly available:} Yes
  \item {\bf Code license:} MIT License
  \item {\bf Data license:} Original datasets (DiffusionDB, MJHQ) follow their respective licenses. Cached images generated by us are released under CC BY 4.0.
  \item {\bf Workflow automation:} Manual execution via provided scripts
\end{itemize}
}

%%%%%%%%%%%%%%%%%%%%%%%%%%%%%%%%%%%%%%%%%%%%%%%%%%%%%%%%%%%%%%%%%%%%%
\subsection{Description}

\subsubsection{How to access}

The artifact code base can be downloaded from \url{https://github.com/stsxxx/MoDM.git}.

\subsubsection{Hardware dependencies}
\begin{itemize}
    \item We conducted our experiments on two systems: 
    a server equipped with an Intel® Xeon® Platinum 8380 CPU and four NVIDIA A40 GPUs (48GB each), 
    and a cluster with 16 nodes, each node containing four AMD MI210 GPUs (64GB each).
    \item The artifact requires a server equipped with 4 GPUs, each having at least 40GB of memory.

\end{itemize}

\subsubsection{Software dependencies}
\begin{itemize}
    \item Ubuntu 22.04 LTS
    \item Python 3.10
    \item CUDA 11.8
    \item PyTorch 2.1.0 compatible with CUDA 11.8
\end{itemize}
\subsubsection{Data sets}
\begin{itemize}
    \item \textbf{DiffusionDB}~\href{https://github.com/poloclub/diffusiondb}{[link]}
    \item \textbf{MJHQ-30k}~\href{https://huggingface.co/datasets/playgroundai/MJHQ-30K}{[link]}
\end{itemize}

\subsubsection{Models}
\begin{itemize}
    \item \textbf{Stable Diffusion 3.5 Large}~\href{https://huggingface.co/stabilityai/stable-diffusion-3.5-large}{[link]}
    \item \textbf{Stable Diffusion XL}~\href{https://huggingface.co/stabilityai/stable-diffusion-xl-base-1.0}{[link]}
    \item \textbf{SANA 1.6B}~\href{https://huggingface.co/Efficient-Large-Model/Sana_1600M_1024px_BF16}{[link]}
\end{itemize}
We use Hugging Face to download and load these models. Note that Stable Diffusion 3.5 Large is a gated model, and requires users to request and obtain access approval before it can be used.
%%%%%%%%%%%%%%%%%%%%%%%%%%%%%%%%%%%%%%%%%%%%%%%%%%%%%%%%%%%%%%%%%%%%%
\subsection{Installation}
Download the MoDM code base from \url{https://github.com/stsxxx/MoDM.git}. Follow the instructions in the README to install all dependencies and to download the dataset metadata, pre-generated cache images, embeddings, and latents.

%%%%%%%%%%%%%%%%%%%%%%%%%%%%%%%%%%%%%%%%%%%%%%%%%%%%%%%%%%%%%%%%%%%%%
\subsection{Experiment Workflow}

\begin{enumerate}
    \item Install all dependencies.
    \item Download and prepare the dataset metadata, as well as all pre-computed cached images and latents.
    \item Run the throughput experiments on MoDM and other baselines.
    \item Compute the image quality metrics.
\end{enumerate}

%%%%%%%%%%%%%%%%%%%%%%%%%%%%%%%%%%%%%%%%%%%%%%%%%%%%%%%%%%%%%%%%%%%%%
\subsection{Evaluation and expected results}

After the experiments complete, all generated images will be saved under the \texttt{images} directory. The throughput and image quality results can be found at the end of each corresponding log file (e.g., \texttt{MoDM\_throughput\_diffusionDB\_sdxl.txt}).

%%%%%%%%%%%%%%%%%%%%%%%%%%%%%%%%%%%%%%%%%%%%%%%%%%%%%%%%%%%%%%%%%%%%%
\subsection{Methodology}

Submission, reviewing and badging methodology:

\begin{itemize}
  \item \url{https://www.acm.org/publications/policies/artifact-review-and-badging-current}
  \item \url{https://cTuning.org/ae}
\end{itemize}

% \bibliographystyle{plain}
% \bibliography{references}

\end{document}